\documentclass[12pt]{article}

\usepackage{epsfig}
\usepackage{amssymb}
\usepackage{amsmath}
\usepackage{color}
\usepackage{hyperref}
\usepackage{bbm}
\usepackage{cite}
\usepackage{geometry}
\makeatletter \@addtoreset{equation}{section} \makeatother

\newcommand{\eq}[1]{\begin{align}#1\end{align}}
\newcommand{\seq}[2]{\begin{subequations}{#1}\begin{align}#2\end{align}\end{subequations}}

\def\la{\left\langle}
\def\ra{\right\rangle}

\def\nn{\nonumber}

\global \long \def \n{\nu}
\global \long \def \m{\mu}

\global \long \def \l{\lambda}
\global \long \def \s{\sigma}

\global \long \def \T{\Theta}
\global \long \def \Tt{{\tilde\Theta}}
\global \long \def \Tb{\bar\Theta}
\global \long \def \Tbt{\tilde{\bar\Theta}}

\global \long \def \t{\theta}
\global \long \def \tt{\tilde\theta}
\global \long \def \tb{\bar\theta}
\global \long \def \tbt{\tilde{\bar\theta}}

\def\Xbfb{{\bar{\mathbf{X}}}}
\def\Xbfbt{{\tilde{\bar{\mathbf{X}}}}}
\def\Xbf{{\mathbf{X}}}
\def\Xbft{{\tilde{\mathbf{X}}}}
\def\Zbf{{\mathbf{Z}}}

\global \long \def \Jm{\mathcal{J}}
\global \long \def \Km{\mathcal{K}}
\global \long \def \Nm{\mathcal{N}}

\global \long \def \Bm{\mathcal{B}}
\global \long \def \Cm{\mathcal{C}}
\global \long \def \Em{\mathcal{E}}

\global \long \def \Dm{\mathcal{D}}
\global \long \def \Qm{\mathcal{Q}}
\global \long \def \Sm{\mathcal{S}}
\global \long \def \Pm{\mathcal{P}}
\global \long \def \Km{\mathcal{K}}

\global \long \def \Rm{\mathcal{R}}
\global \long \def \Am{\mathcal{A}}
\global \long \def \Mm{\mathcal{M}}
\global \long \def \Gm{\mathcal{G}}

\global \long \def \D{\Delta}
\global \long \def \a{\alpha}
\global \long \def \b{\beta}

\global \long \def \d{\delta}
\global \long \def \ad{{\dot\alpha}}
\global \long \def \bd{{\dot\beta}}

\global \long \def \pd{\partial}

\global \long \def \rx{{\rm x}}
\global \long \def \ri{{\rm i}}
\global \long \def \rxt{\tilde{\rm x}}

\def\Nm{{\mathcal{N}}}
\def\Om{{\mathcal{O}}}
\def\Bm{{\mathcal{B}}}
\def\Cm{{\mathcal{C}}}
\def\eps{\epsilon}

\def\pd{\partial}

\def\wh{{\hat\omega}}
\def\wbh{{\hat{\bar\omega}}}
\def\nn{\nonumber}

\newcommand{\bk}[2]{\left\langle \bar{#1}\, #2\right\rangle}
\newcommand{\bkb}[2]{\left\langle #1\, #2\right\rangle}

\newcommand{\tr}{\text{tr}}

\begin{document}

\thispagestyle{empty}
\setcounter{page}{0}

\begin{flushright}\footnotesize
\vspace{0.5cm}
\end{flushright}
\setcounter{footnote}{0}

\begin{center}%
{\Large\textbf{\mathversion{bold}%
Towards general super Casimir equations for $4D$ ${\mathcal N}=1$ SCFTs}\par}

\vspace{15mm}
{\sc 
Israel A. Ram\'irez${}^{a,b}$}\\[5mm]

{\it ${}^{a}$\textit{Departamento de Ciencias Fisicas, Universidad Andres Bello}\\ 
	\textit{Sazie 2212, Santiago de Chile}
}\\[5mm]
{\it ${}^{b}$\textit{Instituut voor Theoretische Fysica, KU Leuven}\\ 
	\textit{Celestijnenlaan 200D, B-3001 Leuven, Belgium}
}\\[5mm]

{\small \texttt{israel.ramirez@kuleuven.be}}\\[20mm]

\begin{abstract}
 Applying the Casimir operator to four-point functions in CFTs allows us to find the conformal blocks for any external operators. In this work, we initiate the program to find the superconformal blocks, using the super Casimir operator, for $4D$ ${\mathcal N}=1$ SCFTs. We begin by finding the most general four-point function with zero $U(1)_R$-charge, including all the possible nilpotent structures allowed by the superconformal algebra. We then study particular cases where some of the operators satisfy shortening conditions. Finally, we obtain the super Casimir equations for four point-functions which contain a chiral and an anti-chiral field. We solve the super Casimir equations by writing the superconformal blocks as a sum of several conformal blocks.
\end{abstract}
\end{center}

\newpage

\setcounter{tocdepth}{2}
\hrule height 0.75pt
\tableofcontents
\vspace{0.8cm}
\hrule height 0.75pt
\vspace{1cm}

\setcounter{tocdepth}{2}

\section{Introduction}

In a general CFTs, it is possible to write the product of two primary operators as a sum of conformal families, given by primary operators and their descendants. This information is encoded in the OPE expansion. The only unknown variable in the OPE is the three-point function coefficient. The fact that conformal symmetry completely fixes the three-point function up to a constant is possible thanks to the absence of conformal invariant structures for any three-point function. Although it is natural to expect a similar statement when supersymmetry enters the game, this is not true in general. The existence of superconformal invariant structures in a generic three-point function was established long ago in the case of four dimensional $\Nm=1$ SCFTs \cite{Park:1997bq,Osborn:1998qu}. Those structures prevent, in general, to write the product of two superconformal primaries as a sum of superconformal families\footnote{For a similar discussion see \cite{Fortin:2011nq}.}. Take, for example, the three-point function of a general scalar multiplet $\Om_1$, with superconformal charges $q$ and $\bar q$, with its conjugated and another scalar long multiplet $\tilde\Om^{(\ell)}$ in an $\Nm=1$ SCFT. Superconformal symmetry fixes the three-point function up to four unrelated constants:
\eq{
	\left \langle \Om_{(p,\bar p)}(z_1) \, \Om_{(\bar p,p)}^{\dagger}(z_2) \, \tilde\Om_{(\D/2,\D/2)}^{(\ell)}(z_3) \right \rangle ={}& \frac{1}{\bk{1}{3}^{\bar p} \bk{3}{1}^p \bk{2}{3}^p \bk{3}{2}^{\bar p}} \frac{1}{\left( \Xbf \cdot \Xbfb\right)^{p+\bar p - (\D-\ell)/2}} \nn \\
	& \times \left[ \l_+^{(1)} t_{1+}^{(\ell)} + \l_-^{(2)} t_{2-}^{(\ell)} + \l_-^{(3)} t_{3-}^{(\ell)} + \l_4^{(4)} t_{4+}^{(\ell)}  \right] \,,\label{long3pf}
}
see \eqref{3pf:OOO} and \eqref{ts:longs}.

Although supersymmetry completely fixes the relation between the superprimary operator of $\Om$ and its super-descendants, the fact that the $\l_i$ coefficients are unrelated implies that there are four distinct conformal families appearing the sOPE of $\Om \times \Om^\dagger$ instead of one single superconformal family. If instead of a long multiplet $\Om$ we study a short-one, for example a chiral multiplet, the shortening conditions will imply relations among the $\l_i$ coefficients \cite{Argyres:2007cn,Poland:2010wg,Rattazzi:2010yc,Ramirez:2016lyk}, but even in those cases it is not always possible to fix all the coefficients, see for example \cite{Fortin:2011nq,Berkooz:2014yda,Khandker:2014mpa,Li:2016chh,Liendo:2015ofa,Manenti:2018xns}.

Furthermore, if one wishes to study the $\l_2$, $\l_3$ and $\l_4$ coefficients in \eqref{long3pf} with, for example, bootstrap techniques \cite{Rattazzi:2008pe}, one has to start with a four-point function which includes all the superconformal invariant terms, otherwise, it would not be possible to perform a complete conformal partial wave expansion. The proliferation of such superconformal invariants for general N-point function in SCFTs has prevented the study of such four-point functions. As we will see, a four dimensional $\Nm=1$ four-point function of scalars with vanishing total $U(1)_R$-charge has 35 nihilpotent structures. Therefore, one has a total of 36, and not just one, independent functions of the conformal invariant cross ratios. This is different from the non-supersymmetric case, where there is only one function of the conformal invariant cross ratios in the four-point function of scalar operators, and it is, as we will see, more closely related to the case where there are spinning correlators \cite{Costa:2011mg}.

A crucial step before we can implement the bootstrap program is to write the four-point function as a sum of conformal blocks (or conformal partial waves.) This has been achieved with great success for CFTs in several dimensions and for several external operators \cite{Dolan:2000ut,Dolan:2003hv,Dolan:2011dv,Costa:2011dw,SimmonsDuffin:2012uy,Echeverri:2015rwa,Echeverri:2016dun,Dymarsky:2017xzb,Cuomo:2017wme}. In supersymmetric theories, this expansion has mostly been obtained at the level of the lowest component of the multiplet, see for example  \cite{Dolan:2004iy,Dolan:2004mu,Poland:2010wg,Rattazzi:2010yc,Fortin:2011nq,Berkooz:2014yda,Khandker:2014mpa,Beem:2014zpa,Bobev:2015jxa,Lemos:2015awa,Doobary:2015gia,Bissi:2015qoa,Li:2016chh,Lemos:2016xke,Li:2017ddj,Bobev:2017jhk,Li:2018mdl}, except in two dimensions where the super Casimir approach was used to find the block expansion for the superconformal invariant structures \cite{Fitzpatrick:2014oza,Lemos:2015awa}.

The aim of this article is to initiate the program for constructing the superconformal partial waves for general four dimensional $\Nm=1$ SCFTs using the super Casimir operator. We begin this program by working out the most general four-point function for scalar with total zero $U(1)_R$-charge. Since such four-point function consists of 36 functions of the supersymetryc cross-ratios in the most general case, we look at four-point functions where at least two of the external operators correspond to short multiplets, in this case, a chiral superfield, $\Phi$, and a current multiplet, $\Jm$. Imposing the shortening conditions, we find that the four-point functions with a chiral and an anti-chiral only depend on three independent functions. Acting with the super Casimir operator on such four-point functions we find a system of coupled eigenvalue equations in the s-channel, while in the t-channel we also find a constraint. In order to solve the super Casimir equations we study the three-point functions, which tell us which operators are being exchange between the lowest components and the first $\Qm$($\bar \Qm$) descendants. This allows us to give an ansatz which solves the super Casimir equations.

This article is structured as follows. In section \ref{Invariants} we find all the four-point function superconformal invariants, which consists on two supersymmetric cross-ratios and 35 nilpotent invariants. This allows us to write down the most general four-point function with zero $U(1)_R$-charge in \eqref{solf}. In section \ref{constrains} we study the cases where some of the operators correspond to a chiral or a current multiplet. The shortening conditions coming from such multiplets imposes several constraints to the four-point functions. In section \ref{casimir} we apply the super Casimir operator on four-point functions where one of the external operators is a chiral and another an anti-chiral and we solve the super Casimir equations by decomposing the superconformal blocks as a sum of conformal blocks. We finish in section \ref{conclusions} with discussions and outline future lines of research.

\section{Constructing the Four-Point Function} \label{Invariants}

The most general four-point function with zero $U(1)_R$-charge is given by
\eq{
	\la \Om_1(z_1) \, \Om_2(z_2) \, \Om_3(z_3) \, \Om_4(z_4) \ra ={}& \frac{\left( \Xbf_{1(23)}^{~2} \right)^a }{\bk{1}{2}^{q_2} \bk{2}{1}^{\bar q_2} \bk{1}{3}^{q_3} \bk{3}{1}^{\bar q_3} \bk{1}{4}^{q_4} \bk{4}{1}^{\bar q_4}} \nn \\ 
	& \qquad \times F(z_1,z_2,z_3,z_4)\,, \label{4pf:general}
}
with
\eq{
	a={}&-q_1+{\bar q}_2+{\bar q}_3+{\bar q}_4\,, \qquad \sum_{i=1}^4(q_i-{\bar q}_i)=0\,.
}
The function $F(z_1,z_2,z_3,z_4)$ is a only a function of the superconformal invariants. In this section, we first briefly review our notation, then we devote ourselves to find the superconformal invariants. We will heavily rely on the results of \cite{Osborn:1998qu}.

\subsection{Notation}

Our notation and conventions will mostly follow \cite{Park:1997bq,Osborn:1998qu} (see also \cite{Wess:320631}). Given the standard supercoordinates $z^A=\left( x^a,\t^\a,\tbt_\ad\right)\in {\mathbb R}^{4|4}$, we can split between the chiral and anti-chiral coordinates,
\eq{
	z_{+}^{A_+}=\left( x_+^a,\t^\a\right)\,,\qquad z_{-}^{A_-}=\left( x_-^a,\tbt_\ad\right)\,,
}
where
\eq{x_{\pm}^a=x^a \pm {\ri}\t\s^a\tb\,.}
These coordinates satisfy $\bar D_\ad z_+=D_\a z_-=0$, where the superderivatives are given by
\eq{
	D_\a=\frac{\pd}{\pd\t^a} + \ri \s_{\a\ad}^m\tb^\ad \pd_m\,,\qquad \bar D_\ad=-\frac{\pd}{\pd \tb^\ad} -\ri \t^\a \s_{\a\ad}^m\pd_m\,.
}
The rule for raising and lowering spinor indices is, as usual, 
\eq{
	\tt_\a=& \epsilon_{\a\b}\t^\b\,, \qquad \tbt_{\ad}=\epsilon_{\ad\bd}\tb^{\bd}\,.
}
Finally, we can write any four-vector as a $2\times 2$ matrix with the help of the Pauli matrices,
\eq{
	\rx_{\a\ad}=x^a \left(\s_a \right)_{\a\ad}\,,
} 
and now the chiral coordinates read
\eq{
	\rx_{\pm}=\rx \mp 2\ri \tt\,\tbt\,,\qquad \rxt_{\pm}=\rxt \pm 2\ri \tb\,\t\,.
}

The supersymmetric version of the distance of two points, $x_{i\,j}=x_i-x_j$ is
\eq{
	\rxt_{\bar i\,j}=&\rxt_{i-} - \rxt_{j+} + 4 \ri \tb_i\,\t_j\,,\qquad \t_{i\,j}=\t_i-\t_j\,, \qquad \tb_{i\,j}=\tb_i-\tb_j. 
}
These coordinates transform under the superconformal group as,
\seq{\label{coordtrans}}{
	\d \rxt_{\bar i\,j}=& \left( \wbh\left( z_{i-} \right) + \bar\s\left( z_{i-} \right) \d \right)\rxt_{\bar i\,j} + \rxt_{\bar i\,j} \left( -\wh\left( z_{j+} \right) + \s\left( z_{j-} \right) \d \right) \,, \label{delx}\\
	\d \t_{i\,j} =& \t_{i\,j} \left( -\wh\left( z_{j+} \right) + \s\left( z_{2+} \right) \d \right) + 2 \left( \bar\s\left( z_{i-} \right) - \s\left( z_{i+} \right) \t_{i\,j} \d \right) -\frac{\ri}{2} \bar\tau \left( z_{i+}\right) \rxt_{\bar i\,j} \,,\\
	\d \tb_{i\,j} =& \left( -\wbh\left( z_{i-} \right) + \bar\s\left( z_{i-} \right) \d \right) \tb_{i\,j} - 2 \left( \bar\s\left( z_{j-} \right) - \s\left( z_{j+} \right) \tb_{i\,j} \d \right) +\frac{\ri}{2} \rxt_{\bar i\,j} \tau \left( z_{j-}\right)\,,
}
being $\wh$, $\wbh$, $\s$, $\bar\s$, $\tau$ and $\bar\tau$ the parameters of the infinitesimal superconformal transformations. Note that the parameters have the following constraints
\eq{
	{\wh_\a}^\a={\wbh^\ad}_\ad=0\,, \qquad D_\a\s=-\frac{1}{3}D_\b {\wh_\a}^\b=\tau_\a \,, \qquad {\bar D}_\a\bar\s=\frac{1}{3}{\bar D}_\b {\wbh_\ad}^\bd={\bar\tau}_\ad\,.
}
It is also useful to define
\eq{
	\left( \rx_{i\,\bar j} \right)_{\a\ad}=- \eps_{\a\b}\eps_{\ad\bd}\left( \rxt_{\bar j \, i} \right)^{\bd\b}\,, \qquad \rx_{i\,\bar j}= \rx_{i+} - \rx_{j-} + 4\, \ri \, \tt_i\,\tbt_j\,,
}
such that,
\eq{
	{\rxt_{\bar i\,j}}^{-1}=\frac{1}{\bk{i}{j}}\rx_{j\,\bar i}\,.,
}
where $\bk{i}{j}=x_{\bar i \, j}^{~2}$.

For $N\ge 3$ points, $z_1,z_2,\cdots\,z_N$ it is convenient to define the supersymmetric coordinate $\Zbf=\left( \Xbf, \T, \Tb\right)$, where the four-vectors $\Xbf$ and its conjugated $\Xbfb$ are given by
\eq{
	\Xbf_{r(s\,t)} =& \frac{\rx_{r\,\bar s}\rxt_{\bar s\,t}\rx_{t\bar r}}{\bk{s}{r}\bk{r}{t}} \,,\qquad \Xbfb_{r(s\,t)}= -\Xbf_{r(t\,s)}\,,  \label{Xs}
}
and the Grassmann variables are defined as
\eq{
	\T_{r(s\,t)}=&\ri \left( \frac{1}{\bk{s}{r}}\tbt_{r\,s}\rxt_{\bar s\,r} - \frac{1}{\bk{t}{r}}\tbt_{t\,s}\rxt_{\bar t\,r}  \right)\,, \qquad \Tb_{r(s\,t)}=\ri \left( \frac{1}{\bk{r}{s}}\rxt_{\bar r\,s} \tt_{r\, s} - \frac{1}{\bk{r}{t}} \rxt_{\bar r\,t} \tt_{r\, t}  \right) \,, \label{Thetas}
}
which transform homogeneously at $z_r$, as can be checked using \eqref{coordtrans}:
\seq{\label{Xtrans}}{
	\d \Xbf_{r(s\,t)}=&\left( \wh\left( z_{r+} \right) - \s\left( z_{r+} \right)\d \right) \Xbf_{r(s\,t)} + \Xbf_{r(s\,t)} \left( -\wbh\left( z_{r-} \right) - \bar\s\left( z_{r-} \right)\d \right)\,, \\
	\d \T_{r(s\,t)}=&\T_{r(s\,t)} \left( -\wh\left( z_{r+} \right) + \s\left( z_{r+} \right)\d -2\,\bar\s \left( z_{r-} \right) \d \right)\,, \\
	\d \Tb_{r(s\,t)}=&\left( \wbh\left( z_{r-} \right) + \bar\s\left( z_{r-} \right)\d - 2\, \s \left( z_{r+} \right) \right) \Tb_{r(s\,t)}\,,
}

As mentioned in the introduction, unlike the non-supersymmetric case, we can construct a superconformal invariant for the three point function
\eq{
	Q=\T_{1(2\,3)}\, {\Xbft_{1(2\,3)}}^{-1}\, \Tb_{1(2\,3)}\,, \label{inv:1}
}
where
\eq{
	{\Xbft_{1(2\,3)}}^{-1}={}& \frac{\Xbf_{1(2\,3)}}{\Xbf_{1(2\,3)}^{~2}} \,.
}
As expected, \eqref{inv:1} vanishes in the bosonic limit. This superconformal invariant was first written by Park \cite{Park:1997bq} with a different notation
\eq{
	I=& \frac{1}{2}\left( \frac{\Xbf_1^{~2}}{\Xbfb_1^{~2}} + \frac{\Xbfb_1^{~2}}{\Xbf_1^{~2}} \right) - 1 = \frac{1}{2} Q^2\,,\\
	J=& \frac{1}{2}\left( \frac{\Xbf_1^{~2}}{\Xbfb_1^{~2}} - \frac{\Xbfb_1^{~2}}{\Xbf_1^{~2}} \right) = - Q \,.
}

As we will see in the next section, $Q$ and $Q^2$ are the only spinless superconformal invariant that can be written in the three-point function.

\subsection{Four-Point Function Invariants}

In flat space we can always go from point $i$ to point $j$ by first passing through a third point $k$: $x_{i\,j}=x_{i\,k}+x_{k\,j}$. A similar statement holds for the supersymmetric version of the distance $x_{i\,\bar j}$,
\eq{
	x_{i\,\bar j}={}&x_{i\,\bar k } + x_{k\,\bar j}- 2\ri\, \t_{i\,k}\s \tb_{j\,k}\,. \label{xs}
}
We can further extend this notion to the vectors given in \eqref{Xs}
\eq{
\Xbf_{r ( s\, t)}=\Xbf_{r (s\, u)} + \Xbf_{r(u\, t)} - 4\, \ri\, \Tt_{r(s\,u)}\, \Tbt_{r(t\,u)}\,, \label{X123}
}
and the Gassmann coordinates \eqref{Thetas},
\eq{
\T_{r ( s\, t)}=\T_{r (s\, u)} + \T_{r(u\, t)}\,, \qquad \Tb_{r ( s\, t)}=\Tb_{r (s\, u)} + \Tb_{r(u\, t)}\,. \label{T123}
}

At first sight, one might think that there are six four-vectors that transform homogeneously at a given point, say $z_1$. But, from \eqref{X123} and using that $\Xbf_{r(s\,s)}=0$, we end up with only two at every point. The same is true for the Grassmann coordinates. In general $N$-point functions, there will be $N-2$ vectors and Grassmann coordinates \cite{Park:1997bq,Park:1998nra,Park:1999pd}. For the moment, we will work on the basis $r(s\,t)=\left\{ 1(2\,3),1(4\,3) \right\}$. Later, we will see that every superconformal invariant can be written in this basis. This also shows that $Q$ and $Q^2$, as given in \eqref{inv:1} and below, are the only three-point function invariants.

In non-supersymmetric theories, there are two conformal invariant functions
\eq{
	u=\frac{\bkb{1}{2} \bkb{3}{4}}{\bkb{1}{3}\bkb{2}{4}} \,, \qquad v=\frac{\bkb{1}{4} \bkb{2}{3}}{\bkb{1}{3}\bkb{2}{4}} \,.
}
The supersymmetric extension of these cross-ratios are given by
\eq{
	u_{r\,s,t\,u}=\frac{\bk{r}{t}\bk{s}{u}}{\bk{r}{u}\bk{s}{t}} \,, \qquad v_{r\,s,t\,u}=\frac{1}{2} \tr \left( \rxt_{\bar r\, t} \rxt_{\bar s\, t}^{~-1} \rxt_{\bar s\, u} \rxt_{\bar r\, u}^{~-1} \right)\,. \label{SCI:uv}
} 
Since their $\theta_i\to 0$ limit is non-vanishing, every other nilpotent superconformal invaritant that we write will be multiplied by an arbitrary function of these $u$ and $v$ terms.

Using \eqref{xs} we can relate any $u_{r\,s,t\,u}$ and $v_{r\,s,t\,u}$ with any other pair $u_{r'\,s',t'\,u'}$ and $v_{r'\,s',t'\,u'}$  plus some $(\t\tb)$ term. The simplest way to do so is by using the $\Xbf$ and $\T$ coordinates. In terms of the $\Xbf$s we have 
\eq{
	\frac{\Xbf_{a}^{~2}}{\Xbf_{b}^{~2}}=\frac{x_{\bar a\, 3}^{~2}\, x_{\bar b\, 1}^{~2}}{x_{\bar a\, 1}^{~2}\, x_{\bar b\, 3}^{~2}}= {u_{a\,b,3\,1}} \, , \qquad \frac{\tr \left( \Xbf_{a} \cdot {\tilde\Xbf}_{b} \right)}{\Xbf_{b}^{~2}}= \frac{\tr \left( \rx_{1\,\bar a}\,\rxt_{\bar a\,3} \, \rx_{3\,\bar b}\, \rxt_{\bar b\,1}  \right)}{x_{\bar a\, 1}^{~2}\, x_{\bar b\, 3}^{~2}}=2 \, v_{b\,a,1\,3}\,,
}
where we have defined,
\eq{
	\Xbf_{a} \equiv \Xbf_{1\left( a\,3 \right)}\,.
}

One might think that terms such as
\eq{
	\frac{\Xbf_a^{~2}}{\Xbfb_b^{~2}}\,,
}
are new superconformal invariants, but from \eqref{Xs} and \eqref{X123} it is easy to see that
\eq{
	\Xbfb_a=\Xbf_a-4\,\ri\,\Tt_a\,\Tbt_a\,, \label{XXb}
}
therefore
\eq{
	\frac{\Xbf_a^{~2}}{\Xbfb_b^{~2}}=\frac{\Xbf_a^{~2}}{\Xbf_b^{~2}}\left(1 - 4\ri \frac{\T_b \,\Xbf_b\, \Tb_b}{\Xbf_b^{~2}} \right)\,,
}
and a similar relation holds for $\tr \left( \Xbfb_a\cdot {\tilde\Xbf}_b \right)/\Xbf_b^2$. If instead of $\Xbf_{1\left( a\, 3 \right)}$, we take $\Xbf_{i\left( a\, j \right)}$ for $i=2,3,4$ we find,
\eq{
	\frac{\Xbf_{i\left( a\, j \right)^{~ 2}}}{\Xbf_{i\left( b\, j \right)^{~ 2}}} = {u_{a\,b,j\,i}} \, , \qquad \frac{\tr \left( \Xbf_{i\left( a\,j \right)} \cdot {\tilde\Xbf}_{i\left( b\,j \right)} \right)}{\Xbf_{i\left( b\,j \right)}^{~2}} = 2 \, v_{b\,a,i\,j}\,. \label{u,v}
}
Terms such as $\Xbf_{i(\cdots)}^{~2}/\Xbf_{j(\cdots)}^{~2}$ are superconformal invariants only if $i=j$. A trace with more than two $\Xbf$ terms, $\tr \left( \Xbf^{2n} \right)$ can always be rewritten as a sum of the cross-ratios by using
\seq{\label{s:identities1}}{
	\left( \s^a\,{\tilde\s}^b\,\s^c + \s^c\,{\tilde\s}^b\,\s^a  \right)_{\a\,\b} = & 2 \left( \eta^{a\,c}\s^b -\eta^{a\,b}\s^c - \eta^{b\,c}\s^a\right)_{\a\,\b}\,, \\
	\left( \s^a {\tilde\s}^b + \s^a {\tilde\s}^b \right)_{\a}^{~\b} = & -2 \eta^{a\,b}\d_\a^{~\b}\,,
}
since we only have two $\Xbf$-terms that can appear in such expression.

We can now easily relate $u_{ab,ij}$ with $u_{ij,ab}$
\eq{
	u_{ij,ab}={}& \frac{\rx_{\bar i a}\rx_{\bar j b}}{\rx_{\bar i b}\rx_{\bar j a}}=\frac{\Xbfb_a^{~2}}{\Xbfb_b^{~2}}=\frac{\Xbf_a^{~2}}{\Xbf_b^{~2}}+\Om\left( \T\Tb \right) = u_{ab,ij} + \cdots\,,
}
where $\cdots$ means terms proportional to $\t\tb$. A similar relations hold for the $v$'s defined in \eqref{u,v}. Finally, note that,
\eq{
	\frac{\bk{1}{3}\bk{2}{4}}{\bk{1}{4}\bk{2}{3}}=\frac{\Xbf_{1(42)}^{~2}}{\Xbf_{1(32)}^{~2}}=\frac{(\Xbf_{1(23)}-\Xbf_{1(43)})^{~2}}{\Xbf_{1(32)}^{~2}}+\Om\left( \T\Tb \right)=1-v_{24,13}+u_{42,31}+\cdots\,.
}
We can perform a similar exercise for any other combination. This shows that any arbitrary function $f\left( u_{ab,cd},v_{ij,kl} \right)$ can be written as $f\left( u_{24,31},v_{42,13} \right)$ up to fermionic terms, as already mentioned.

Therefore, our first term of the $\T$-expansion of $F(z_1,z_2,z_3,z_4)$ is just an arbitrary function of the cross-ratios, as expected,
\eq{
	F(z_1,z_2,z_3,z_4)=A\left( u,v \right) + \Om \left( \T\,\Tb \right)\,.
}

At the first $\T$ level, the only scalar we can construct with zero $U(1)_r$-charge are
\eq{
	\frac{\T_{i\left( a\,j \right)}\,\Xbf_{i\left( b\,j \right)}^{~2n+1}\,\Tb_{i\left( c\,j \right)}}{\Xbf_{i\left( d\,j \right)}^{~n+1}} \,,
}
for $i=1,2,3,4$. Using \eqref{s:identities1}, we can write any of those terms as a function of
\eq{
	\T_{i\left( a\,j \right)}\, \Xbft_{i\left( b\,j \right)}^{~-1} \, \Tb_{i\left( c\,j \right)}\, \label{TXT:1}
}
times an arbitrary function of the cross-ratios. We now show that we can write any term in \eqref{TXT:1} as a function of,
\eq{
	\left( a\,b\,c \right)=\T_a\, \Xbft_b^{-1} \, \Tb_c\,, \label{txt:def}
}
times an arbitrary function of the cross-ratios up to $\Om\left( \T^2\,\Tb^2 \right)$. In order to do so, we need to generalize the relation (2.52) and (2.53) in \cite{Osborn:1998qu}:
\seq{\label{rotations}}{
	\rxt_{\bar j\,i} \, \Xbf_{i\left( j\,k \right)}\, \rxt_{\bar i\,j} = & \frac{1}{\Xbfb_{j \left( k\,i \right)}^{~2}} \Xbfbt_{k \left( j\,i \right)} \,, \\
	\frac{x_{\bar i\,j}^{~2}}{x_{\bar j\,i}^{~2}} \T_{i\left( j\,k \right)}\, \rx_{ i\,\bar j} = & \frac{1}{\Xbf_{j\left( k\,i \right)}^{~2}} \, \T_{j\left( k\,i \right)} \,\Xbf_{j\left( k\,i \right)}\,,\\
	\frac{x_{\bar j\,i}^{~2}}{x_{\bar i\,j}^{~2}} \rx_{j\,\bar i} \, \Tb_{i\left( j\,k \right)}  = & \frac{1}{\Xbfb_{j\left( k\,i \right)}^{~2}} \, \Xbfb_{j\left( k\,i \right)}\, \Tb_{j\left( k\,i \right)}\,,
}
which can be easily verified using
\eq{
	\rxt_{\bar a\, b }= \rxt_{\bar a\, c} + \rxt_{\bar c\, b} + 4\,\ri\,\tb_{a\,c}\,\t_{b\,c}\,.
}
For example,
\eq{
	\T_{1\left( a\, 3 \right)}\, \Xbft_{1\left( b\, 3 \right)}^{~-1}\, \Tb_{1\left( c\, 3 \right)} = & \, \T_{1\left( 3\, a \right)}\, \Xbft_{1\left( 3\, b \right)}^{~-1}\, \Tb_{1\left( 3\, c \right)} + \cdots \\
	= & \, \frac{1}{\Xbf_{3\left( a\,1 \right)}\Xbfb_{3\left( c\,1 \right)}} \T_{3\left( a\,1 \right)}\, \Xbf_{3\left( a\,1 \right)} \, \Xbfbt_{3\left( b\,1 \right)} \, \Xbfb_{3\left( c\,1 \right)} \, \Tb_{3\left( c\,1 \right)} +\cdots \\
	= & \, \frac{1}{\Xbf_{3\left( a\,1 \right)}\Xbf_{3\left( c\,1 \right)}} \T_{3\left( a\,1 \right)}\, \Xbf_{3\left( a\,1 \right)} \, \Xbft_{3\left( b\,1 \right)} \, \Xbf_{3\left( c\,1 \right)} \, \Tb_{3\left( c\,1 \right)} +\cdots \,, \label{1to3:1}
}
where the dots means higher powers in $\T\,\Tb$. We have repeatedly used \eqref{Xs}, \eqref{Thetas}, \eqref{X123}, \eqref{T123} and \eqref{XXb}. As discussed before, for the ${3(i\,1)}$ basis we only need $i=2,4$. Therefore, we can have either $a=b$, $b=c$ or $a=c$ in \eqref{1to3:1}. Using \eqref{s:identities1} we find that,
\seq{\label{1to3:2}}{
	\left( a\,a\,b \right) = & \frac{1}{\Xbf_{3\left( b\,1 \right)}^{~2}} \T_{3\left( a\,1 \right)}\, \Xbf_{3\left( a\,1 \right)}\, \Tb_{3\left( b\,1 \right)} +\cdots = u_{b\,a,1\,3}\,\T_{3\left( a\,1 \right)}\, \Xbft_{3\left( a\,1 \right)}^{-1}\, \Tb_{3\left( b\,1 \right)} +\cdots\,,\\
	\left( a\,b\,b \right) = & \frac{1}{\Xbf_{3\left( a\,1 \right)}^{~2}} \T_{3\left( a\,1 \right)}\, \Xbf_{3\left( b\,1 \right)}\, \Tb_{3\left( b\,1 \right)} +\cdots = u_{a\,b,1\,3}\,\T_{3\left( a\,1 \right)}\, \Xbft_{3\left( b\,1 \right)}^{-1}\, \Tb_{3\left( b\,1 \right)} +\cdots\,,\\
	\left( a\,b\,a \right) = & \frac{1}{\Xbf_{3\left( a\,1 \right)}^{~2} } \T_{3\left( a\,1 \right)}\, \Xbf_{3\left( b\,1 \right)}\, \Tb_{3\left( a\,1 \right)} + \frac{\tr\left( \Xbf_{3\left( a\,1 \right)\cdot \Xbft_{3\left( b\,1 \right)}} \right)}{\Xbf_{3\left( a\,1 \right)}^{~2}} \T_{3\left( a\,1 \right)}\, \Xbft_{3\left( a\,1 \right)}^{~-1}\, \Tb_{3\left( a\,1 \right)} +\cdots \nn\\ 
	=& u_{a\,b,1\,3}\,\T_{3\left( a\,1 \right)}\, \Xbft_{3\left( b\,1 \right)}^{-1}\, \Tb_{3\left( b\,1 \right)} + 2 v_{a\,b,3\,1} \T_{3\left( a\,1 \right)}\, \Xbft_{3\left( a\,1 \right)}^{-1}\, \Tb_{3\left( a\,1 \right)} \cdots \,,
}
where $\left( a\,b\,c \right)$ was defined in \eqref{txt:def}. This shows that we can always change the basis by using \eqref{Xs} and \eqref{Thetas}. For example, we can write any $\Xbf_{2(a\,b)}$ in terms of $\Xbf_{2\left( 1\,3 \right)}$ and $\Xbf_{2\left( 1\,4 \right)}$, and since we can change our original basis from $\Xbf_{1\left( a\,3 \right)}$ to $\Xbf_{1\left( b\,2 \right)}$ we can have a result similar to \eqref{1to3:2}. Thus, the $\left( a\,b\,c \right)$ terms in \eqref{txt:def} are a basis for all the possible superconformal invariant at the first $\T\,\Tb$ level.

Are all the eight different $\left( a\,b\,c \right)$ terms independents? In order to answer this, we perform a suitable superconformal transformation and set $z_1^A=\left( 0,0,0 \right)$ and $z_3^A=\left( \infty,0,0 \right)$. In this basis, we find that,
\eq{
	\left( a\,b\,c \right)=\frac{\t_c\, \rx_{c+}\,\rxt_{b-}\,\rx_{a-}\tb_a}{x_{a-}^{~2}x_{c+}^{~2}}\,,
}
which indeed shows that the eight $\left( a\,b\,c \right)$ terms are linearly independent. Thus, our $F(z_1,z_2,z_3,z_4)$ function now reads
\eq{
F(z_1,z_2,z_3,z_4) = & A + \sum_{a,b,c=2,4} B_{a,b,c}\left(a\,b\,c\right)+\Om \left( \T^2 \Tb^2\right)\,.
}

At the $\T^2\,\Tb^2$-level, we can write the following superconformal invariants
\seq{}{
	\frac{\left( \T_a\,\Tt_b \right) \left( \Tbt_c\,\Tb_d \right)}{\Xbf_e^{~2}} \,,&\qquad \frac{\left( \T_a\,\Tt_b \right) \left( \Tbt_c\, \Xbft_{d}\,\Xbf_e\Tb_f \right)}{\Xbf_g^{~2}}\,,\qquad \frac{\left( \T_a\,\Xbf_b\,\Xbft_c\,\Tt_d \right) \left( \Tbt_e\,\Tb_f \right)}{\Xbf_g^{~2}}\,, \nn \\
	\left( a\,b\,c \right) \left( d\,e\,f \right)\,, &\qquad \frac{\left( \T_a\,\Xbf_b\,\Xbft_c\,\Tt_d \right) \left( \Tbt_e\,\Xbft_f\,\Xbf_g\,\Tb_h \right)}{\Xbf_i^{~2}} \,.
}
Using the Fierz identities, we can relate any of these superconformal invariants with $\left( a\,b\,c \right) \left( d\,e\,f \right)$. First, we can check that
\eq{
	\left( \T_a\,\Tt_b \right) \left( \Tbt_c\,\Tb_d \right) =\frac{1}{\tr\left( x\cdot{\tilde y} \right) } \left( \left( \T_a \, x\, \Tb_c \right)\left( \T_b\,y\,\Tb_d \right) + (a\leftrightarrow b) + (c \leftrightarrow d) \right)\,,
}
for any two vector $x$ and $y$, which we can choose as $\Xbf_{1\left( 2\,3 \right)}$ and $\Xbf_{1\left( 4\,3 \right)}$. Thus, these terms reduce to the $\left( a\,b\,c \right) \left( d\,e\,f \right)$ type of terms.

Note that we can write any $\left( \T_a\,\Xbf_b\,\Xbft_c\,\Tt_d \right)$ as a function of $\left( \T_a\,\Tt_d \right)$ and $\left( \T_a\,\Xbf_2\,\Xbft_4\,\Tt_d \right)$ using \eqref{s:identities1}, and similarly for the $\Tb$ situation. Now we can check that
\seq{}{
	\left( \T_a\,\Xbf_2\,\Xbft_4\,\Tt_b \right)\left( \Tbt_c \,\Tb_d \right) = & \left[ \left( a\,2\,c  \right)\left( b\,4\,d \right)+\left( a\,2\,d \right)\left( b\,4\,c \right)\right] \left( \Xbf_2^{~2}\,\Xbf_4^{~2} \right)\,, \\
	\left( \T_a\,\Tt_b \right)\left( \Tbt_c \,\Xbft_2\,\Xbft_4\,\Tb_d \right) = & \left[ \left( a\,2\,c  \right)\left( b\,4\,d \right)+\left( a\,4\,d \right)\left( b\,2\,c \right)\right] \left( \Xbf_2^{~2}\,\Xbf_4^{~2} \right)\,.  
}

The $\left( \T_a\,\Xbf_b\,\Xbft_c\,\Tt_d \right) \left( \Tbt_e\,\Xbft_f\,\Xbf_g\,\Tb_h \right)$ terms are trivially reduced to the previous cases except for $a=e=2$ and $d=h=4$. In those cases, one can finally show that
\eq{
	\left(\frac{ \T_2\,\Xbf_2\,\Xbft_4\,\Tt_4 }{\Xbf_2^{~2}}\right) \left( \frac{ \Tbt_2\,\Xbft_2\,\Xbf_4\,\Tb_4 }{\Xbf_4^{~2}} \right) ={} & \Xbf_2^{~2} \left( 2\,2\,4 \right) \left( 4\,2\,2 \right)+2\frac{\tr\left( \Xbf_2\cdot \Xbft_4 \right)}{\Xbf_2^{~2}\,\Xbf_4^{~2}} \left( 2\,2\,2 \right)\left( 4\,4\,4 \right)\nn\\
	& - \Xbf_4^{~2} \left( 2\,4\,2 \right)\left( 4\,4\,4 \right)\,.
}
Thus, at the second order in $\T\,\Tb$ we can write everything as a function of the $\left( a\,b\,c \right)\left( d\,e\,f \right)$ terms. But not all of those 64 terms are linearly independent. Everything can be written in terms of 18 independent terms. We can choose the following basis 
\seq{\label{sci:tt}}{
	&\left\{ \left( 2\,2\,2 \right)^2 \right\} \,, \\
	&\left\{ \left( 2\,4\,4 \right)\left( 2\,2\,4 \right) \right\} \,, \\
	&\left\{ \left( 2\,4\,2 \right)\left( 2\,2\,4 \right),\left( 2\,2\,2 \right)\left( 2\,4\,4 \right) \right\} \,, \\
	&\left\{ \left( 2\,2\,2 \right)\left( 4\,4\,2 \right),\left( 2\,4\,2 \right)\left( 4\,2\,2 \right) \right\} \,, \\
	&\left\{\left( 2\,4\,4 \right)\left( 4\,2\,2 \right),\left( 2\,4\,2 \right)\left( 4\,4\,4 \right),\left( 2\,2\,2 \right)\left( 4\,2\,4 \right), \left( 2\,2\,2 \right)\left( 4\,4\,4 \right), \left( 2\,2\,4 \right)\left( 4\,4\,2 \right),\right. \nn\\
	&\left.\left( 2\,4\,2 \right)\left( 4\,2\,4 \right)\right\}\,, \\
	&\left\{ \left( 2\,4\,4 \right)\left( 4\,2\,4 \right),\left( 2\,2\,4 \right)\left( 4\,4\,4 \right) \right\} \,,\\
	&\left\{ \left( 4\,2\,2 \right)\left( 4\,4\,4 \right), \left( 4\,2\,4 \right)\left( 4\,4\,2 \right) \right\} \,,\\
	&\left\{ \left( 4\,4\,2 \right)\left( 4\,2\,2 \right) \right\} \,, \\
	&\left\{ \left( 4\,4\,4 \right)\left( 4\,4\,4 \right) \right\}\,.
}
We can write any of the remaining 45 terms as a linear combination of \eqref{sci:tt}. This can be verified by with the help of the following relations
\seq{}{
	\frac{\left( a\,m\,b \right)\left( a\,n\,b \right)}{\left( \Xbf_m\cdot \Xbf_n \right)} = & \frac{\left( a\,p\,b \right)\left( a\,q\,b \right)}{\left( \Xbf_p\cdot \Xbf_q \right)}\,,\\
	2\frac{\left( a\,m\,b \right)\left( a\,m\,c \right)}{\Xbf_m^{~2}} = & \frac{\left( a\,m\,b \right)\left( a\,n\,c \right)+\left( a\,n\,b \right)\left(a\,m\,c  \right)}{\left( \Xbf_n\cdot\Xbf_m \right)} \,, \\
	2\frac{\left( a\,m\,b \right)\left( c\,m\,b \right)}{\Xbf_m^{~2}} = & \frac{\left( a\,m\,b \right)\left( c\,n\,b \right)+\left( a\,n\,b \right)\left(c\,m\,b  \right)}{\left( \Xbf_n\cdot\Xbf_m \right)} \,, \\
	\frac{\left( a\,m\,b \right)\left( c\,m\,d \right)+\left( a\,m\,d \right)\left( c\,m\,b \right)}{\Xbf_m^{~2}} = & \frac{\left( a\,n\,b \right)\left( c\,n\,d \right)+\left( a\,n\,c \right)\left( c\,n\,b \right)}{\Xbf_n^{~2}} \,.
}

From what we learned from the previous case, we can rotate our basis and reach any $\left\{ i\left(j\,k \right) \right\}$ basis. Thus, we have exhausted all the possibilities at order $\T^2\,\Tb^2$. Therefore, our $f$ function is,
\eq{
	F(z_1,z_2,z_3,z_4) ={} & A + \sum_{a,b,c=2,4} B_{a,b,c}\left(a\,b\,c\right) + \sum_{a_i=2,4} C_{a_i}\left(a_1\,a_2\,a_3\right) \left(a_4\,a_5\,a_6 \right)+\Om(\T^3\Tb^3)\,,
}
where the $C_{a_i}$ are arbitrary function of the cross-ratios and non-zero only for the cases listed in \eqref{sci:tt}.

The $\T^3\,\Tb^3$ case is now almost straightforward. From the previous discussion, we know that we can only have structures of the type
\eq{
	\left( a_1\,b_1\,c_1 \right)\left( a_2\,b_2\,c_2 \right)\left( a_3\,b_3\,c_3 \right)\,.
}
The only question is, which structures are independent. Since our basis $1\left( a\,3 \right)$ can only take two values, $a=1,2$, we have only two types of terms
\seq{}{
	\left( \psi\,\Xbf\, {\bar \chi}  \right)\left( \psi\,{\mathbf Y} \, \bar \chi  \right) \left( \eta\,{\mathbf Z}\, \bar\xi \right)=& \frac{1}{4} \left( \psi^2 \right)\left( {\bar \chi}^2 \right) \left( \eta\,\mathbf Z\,\bar\xi  \right) \left( \tr\,\Xbft\,\mathbf Y \right) \,, \\
	\left( \psi\,\Xbf\, {\bar \chi}  \right)\left( \eta\,{\mathbf Y} \, \bar \chi  \right) \left( \psi\,{\mathbf Z}\, \bar\xi \right)=& -\frac{1}{4} \left( \psi^2 \right)\left( {\bar \chi}^2 \right) \left( \eta\,\mathbf Y\,\Xbft\,\mathbf Z\,\bar\xi  \right) \,.
}
Therefore, since
\eq{
	\Xbf_a\,\Xbft_b\,\Xbf_c=& \left\{ \begin{matrix}
					\Xbf_a \Xbf_b^{~2} & b=c \\
					\left[ \Xbf_a \left( \Xbf_b\cdot\Xbf_a \right)+\Xbf_b\left( \Xbf_a^{~2} \right) \right] & a=c 
	\end{matrix}\right.\,,
}
the only terms that we can have are
\seq{}{
	\left( \T_2^{~2} \right)\left( \Tb_2^{~2} \right)\left( 4\,a\,4 \right)\,,\\
	\left( \T_2^{~2} \right)\left( \Tb_4^{~2} \right)\left( 4\,a\,2 \right)\,,\\
	\left( \T_4^{~2} \right)\left( \Tb_2^{~2} \right)\left( 2\,a\,4 \right)\,,\\
	\left( \T_4^{~2} \right)\left( \Tb_4^{~2} \right)\left( 2\,a\,2 \right)\,,
}
which can be arranged as,
\seq{}{
	\left( 2\,2\,2 \right)^2\left( 4\,a\,4 \right)\,, \qquad \left( 2\,2\,2 \right)\left( 4\,4\,4 \right)\left( 4\,a\,2 \right)\,,\\
	\left( 4\,4\,4 \right)^2\left( 2\,a\,2 \right)\,, \qquad \left( 2\,2\,2 \right)\left( 4\,4\,4 \right)\left( 2\,a\,4 \right)\,.
}

Finally, now it is easy to see that the $\T^4\,\Tb^4$ term is unique,
\eq{
	\left( 2\,2\,2 \right)^2\left( 4\,4\,4 \right)^2,
}
and higher order in $\T\,\Tb$ vanish. Thus, the most general solution for $F(z_1,z_2,z_3,z_4)=F\left(\Zbf_a\right)$ is given by
\eq{
	F\left(\Zbf_a\right) ={} & \Am_1 + 4\ri \Bm_1 \left(222\right)+ 4\ri \Bm_2 \left(242\right)+ 4\ri \Bm_3 \left(224\right) + 4\ri \Bm_4 \left(244\right) + 4\ri \Bm_5 \left(422\right) \nn \\ & + 4\ri \Bm_6 \left(442\right) + 4\ri \Bm_7 \left(424\right) + 4\ri \Bm_8 \left(444\right) + \Cm_1 \left(222\right)\left(222\right) +\Cm_2 \left(244\right)\left(224\right) \nn \\ & + \Cm_3 \left(242\right)\left(224\right) + \Cm_4 \left(222\right)\left(244\right)  + \Cm_5 \left(222\right)\left(442\right) + \Cm_6 \left(242\right)\left(422\right) \nn \\ & + \Cm_7 \left(244\right)\left(422\right) + \Cm_8 \left(242\right)\left(444\right) + \Cm_9 \left(222\right)\left(424\right) + \Cm_{10} \left(222\right)\left(444\right) \nn \\ & + \Cm_{11} \left(224\right)\left(442\right) + \Cm_{12} \left(242\right)\left(424\right) + \Cm_{13} \left(244\right)\left(424\right) + \Cm_{14} \left(224\right)\left(444\right) \nn \\ & + \Cm_{15} \left(422\right)\left(444\right) + \Cm_{16} \left(424\right)\left(442\right) + \Cm_{17} \left(442\right)\left(422\right) + \Cm_{18} \left(444\right)\left(444\right) \nn \\ & + 4\ri \Dm_{1} \left(222\right)\left( 222 \right)\left(424\right) + 4\ri \Dm_{2} \left(222\right)\left( 222 \right)\left(444\right) + 4\ri \Dm_{3} \left(222\right)\left(444\right)\left(422\right) \nn \\ & + 4\ri \Dm_{4} \left(222\right)\left(444\right)\left(442\right) + 4\ri \Dm_{5} \left(444\right)\left(444\right)\left(222\right) +  4\ri \Dm_{6} \left(444\right)\left(444\right)\left(242\right) \nn \\ & + 4\ri \Dm_7 \left(222\right)\left(444\right)\left(224\right) + 4\ri \Dm_8 \left(222\right)\left(444\right)\left(244\right) + \Em_{1} \left(222\right)^2\left(444\right)^2\,. \label{solf}
}

The form of \eqref{solf} allows to change $\Xbf_{1(23)}$ by $\Xbf_{1(43)}$ in \eqref{4pf:general}, and this will only imply a shift in the $f(u,v)$ functions. Furthermore, we can also make a change of basis, as we will later do, by following the same procedure as in \eqref{1to3:2}. This will allow us to impose constrains at any point of the four point function.

\section{Imposing Constraints to the Four-Point Function} \label{constrains}

Plugging \eqref{solf} into \eqref{4pf:general} give us the most general four-point function, but, as we already know from experience with the three-point functions, shortening conditions impose relations among the different structures. In this section we will study such constraints in the case of the four-point function. We will study three systems of increasing complexity. The simplest system consist only on chiral and anti-chiral fields
\eq{
	\left\langle \Phi(z_1)\,\bar \Phi(z_2)\,\Phi(z_3)\,\bar\Phi(z_4) \right\rangle \,.
}
Such fields satisfy
\eq{
	{\bar D}_\ad \Phi={D}_\a \bar\Phi=0 \,. \label{ChiralConstrain}
}
We are also interested in a mixed system of chirals and current-multiplets,
\eq{
	\left\langle J(z_1)\,J(z_2)\,\Phi(z_3)\,\bar\Phi(z_4) \right\rangle \,,
}
where the current-multiplets satisfy
\eq{
	{D^2}J= {\bar D}^2 J=0\,. \label{CurrentConstrain}
}
Finally, we will study  a system with only conserved currents
\eq{
	\left\langle J(z_1)\,J(z_2)\,J(z_3)\,J(z_4) \right\rangle \,.
}

\subsection{From Superderivatives to Superconformal Covariant Derivatives}

In order to understand the constraints \eqref{ChiralConstrain} and \eqref{CurrentConstrain} we first need to see how the super-derivatives acts on $\Zbf_{a(bc)}$ coordinates. It is not hard to generalize the action of the super-derivatives on the $\Zbf_{a(bc)}$ given in \cite{Osborn:1998qu}:
\seq{\label{dtoD}}{
	D_{j\a} \, \Xbf_{i\left( j\,k \right)}^a ={}& 0 \,, &
	\bar D_{j\ad} \, \Xbf_{i\left( j\,k \right)}^a ={}& -\ri \frac{1}{\bk{i}{j}} \left( \rx_{i\,\bar j}\right)_{\a\,\ad} \eps^{\a\,\b} \left( -2\ri \s^a \Tb_{i\left( j\,k \right)} \right)_\a\,, \\
	D_{j\a} \, \T_{i\left( j\,k \right)}^\b ={}& 0 \,, &
	\bar D_{j\ad} \, \T_{i\left( j\,k \right)}^\b ={}&  -\ri \frac{1}{\bk{i}{j}} \left( \rx_{i\,\bar j}\right)_{\a\,\ad} \eps^{\a\,\b}\,, \\
	D_{j\a} \, \Tb_{i\left( j\,k \right)}^\bd ={}& \ri \frac{1}{\bk{j}{i}}\left( \rx_{j\,\bar i}\right)_{\a\,\ad}\eps^{\ad\,\bd}\,, &
	\bar D_{j\ad} \, \Tb_{i\left( j\,k \right)}^\bd ={}&  0\,, \\
	\bar D_{k\ad} \, \Xbf_{i\left( j\,k \right)}^a ={}& 0 \,, &
	D_{k\a} \, \Xbf_{i\left( j\,k \right)}^a ={}&  -\ri \frac{1}{\bk{k}{i}}\left( \rx_{i\,\bar k} \right)_{\a\,\ad}\eps^{\ad\,\bd} \left( 2\ri\,\T_{i\left( j\,k \right)} \s^a \right)_{\bd}\,, \\
	\bar D_{k\ad} \, \T_{i\left( j\,k \right)}^\b ={}& \ri \frac{1}{\bk{i}{k}}\left( \rx_{i\,\bar k} \right)_{\a\,\ad} \eps^{\a\b} \,, &
	D_{k\a} \, \T_{i\left( j\,k \right)}^\b ={}&  0\,, \\
	\bar D_{k\ad} \, \Tb_{i\left( j\,k \right)}^\bd ={}& 0 \,, &
	D_{k\a} \, \Tb_{i\left( j\,k \right)}^\bd ={}& -\ri \frac{1}{\bk{k}{i}}\left( \rx_{i\,\bar k} \right)_{\a\,\ad}\eps^{\ad\,\bd}  \,.
}
By inspecting \eqref{dtoD} it is natural to define the superconformal covariant derivatives\footnote{An $\Nm=2$ version of these operators is given in \cite{Kuzenko:1999pi}.}
\seq{\label{SCCD}}{
	\Dm_{\a}={}& \frac{\pd}{\pd \T^\a_{i\left( j\,k \right)}}-2\ri \left( \s^a\Tb_{i\left( j\,k \right)} \right)_\a\frac{\pd}{\pd \Xbf_{i\left( j\,k \right)}}\,, & 
	\bar \Dm_{\ad}={}& - \frac{\pd}{\pd \Tb^\ad_{i\left( j\,k \right)}}\,,\\
	\bar \Qm_{\ad}={}& \frac{\pd}{\pd \Tb^\ad_{i\left( j\,k \right)}}+2\ri \left( \T_{i\left( j\,k \right)}\s^a \right)_\a\frac{\pd}{\pd \Xbf_{i\left( j\,k \right)}}\,, & 
	\Qm_{\a}={}& -\frac{\pd}{\pd \T^\a_{i\left( j\,k \right)}}\,.
}

As we will see later, the shortening conditions \eqref{ChiralConstrain} and \eqref{CurrentConstrain} acting on the four-point function take a simpler form as a superconformal covariant derivative \eqref{SCCD} acting on $F(\Zbf_a)$.

\subsection{
\texorpdfstring{$\left\langle \Phi \,\bar \Phi \,\Phi \,\bar\Phi \right\rangle$}{phi phidagger phi phidagger}}
The simplest four-point function that we can solve is
\eq{
	\left\langle \Phi(z_1)\,\bar \Phi(z_2)\,\Phi(z_3)\,\bar\Phi(z_4) \right\rangle ={}& \frac{1}{\bk{2}{1}^{q} \bk{4}{3}^{q}} F\left( \Zbf_a\right)\,, \label{4pf:pbppbp}
}
where $F\left( \Zbf_a\right)$ is given by \eqref{solf}.

The most straightforward way to solve \eqref{4pf:pbppbp} is by using a system which is explicitly chiral at $z_{1,3}$ and anti-chiral at $z_{2,4}$. A basis with such behavior is 
\eq{
	\Xbf_{a}={} & \Xbf_{1(a3)} \,, & \T_{a}={}& \T_{1(a3)}\,, & \Tb_{a}={}& \Tb_{1(a3)}\,.
}
Although the cross-ratios $U$ and $V$,
\eq{
	U={} & \frac{\Xbf_2^{~2}}{\Xbf_4^{~2}}=\frac{\bk{2}{3} \bk{4}{1}}{\bk{4}{3} \bk{2}{1}}	\,, & V={}&\frac{\left( \Xbf_2 \cdot \Xbf_4 \right)}{\Xbf_4^{~2}} = -\frac{1}{2} \tr\left( \frac{\rx_{1\bar2}\rxt_{\bar2 3}\rx_{3\bar 4}\rxt_{\bar 4 1}}{\bk{2}{1} \bk{4}{3}} \right)\,.
}
have the right chirality for every $z_i$, they do not reduce to the standard cross-ratios in the bosonic case. Thus, we will use
\eq{
	u={}&\frac{1}{1+U-2V} \,, \qquad v=\frac{U}{1+U-2V}\,, \label{cross-ratios:good}
}
which in the bosonic limit are
\eq{
	u|_{\t,\tb\to0}=\frac{\bkb{1}{2} \bkb{3}{4}}{\bkb{1}{3} \bkb{2}{4}} \,, \qquad v|_{\t,\tb\to0}=\frac{\bkb{1}{4} \bkb{2}{3}}{\bkb{1}{3} \bkb{2}{4}} \,,
}
as desired.

Using \eqref{dtoD}, we can write the constraints to the four-point function as differential equations acting on the $F$ function in \eqref{4pf:pbppbp}. Explicitly,
\seq{\label{chiral-constrants}}{
	D_{4\a}\bar\Phi (z_4)={}&0 \qquad\Rightarrow  \qquad \Qm_{4\a}F(\Zbf)={}0\,, \\
	D_{2\a}\bar\Phi(z_2)={}&0 \qquad\Rightarrow  \qquad \Qm_{2\a}F(\Zbf)={}0\,, \\
	{\bar D}_{3\ad}\Phi (z_3)={}&0 \qquad\Rightarrow  \qquad ({\bar\Dm}_{2\ad}+{\bar\Dm}_{4\ad})F(\Zbf)={}0\,,
}

It is simple to see that the first two constraints totally fix $F(\Zbf_a)$ setting all the $\Bm$s, $\Cm$s and $\Em_1$ functions to vanish. Therefore,
\eq{
	\left\langle \Phi(z_1)\,\bar \Phi(z_2)\,\Phi(z_3)\,\bar\Phi(z_4) \right\rangle ={}& \frac{1}{\bk{2}{1}^{q} \bk{4}{3}^{q}} \Am_1(u,v)\,, \label{4pf:pbppbp_sol}
}
with $u$ and $v$ given above.

Since we never used the chirality property of the $\Phi$s, one might wonder what happens when the chiral fields are replaced by arbitrary scalar multiplets. Taking the scalar fields $\Om_1$ and $\Om_2$ with $U(1)_R$-charges $(q_1,{\bar q}_1)$ and $(q_2,{\bar q}_2)$ respectively, we find
\eq{
	\left\langle \Om_1(z_1)\,\bar \Phi(z_2)\,\Om_2(z_3)\,\bar\Phi(z_4) \right\rangle ={}& \frac{\left( \Xbf_2^{~2} \right) ^{{\bar q}_1-q_3}}{\bk{2}{1}^{\bar q} \bk{4}{1}^{\bar q} \bk{1}{3}^{q_3} \bk{3}{1}^{{\bar q}_3}} \Am_1(u,v)\,, \label{4pf:ObpPbp_sol}
}
with $q_1 + q_2 = {\bar q}_1 + {\bar q}_2 + 2{\bar q}$. Note that the fact that \eqref{4pf:pbppbp_sol} and \eqref{4pf:ObpPbp_sol} do not depend on any nilpotent structure simplified the computation of the super Casimir acting on the four-point function \cite{Fitzpatrick:2014oza,Bobev:2015jxa}.

\subsection{
\texorpdfstring{$\left\langle \Jm \, \Jm \, \Phi \, {\bar\Phi} \right\rangle$}{J J phi phidagger}}

More interesting is the case where the first two fields are replaced by two current-multiplets,
\eq{
	\left\langle \Jm(z_1) \, \Jm(z_2) \, \Phi(z_3) \, {\bar\Phi}(z_4) \right\rangle	=&{} \frac{1}{ \bk{1}{2} \bk{2}{1} \bk{4}{3}^{q}} F\left( \Zbf_a\right) \,. \label{4pf:jjpbp}
}

Again, we first set, as in the previous computation, a basis which is explicitly chiral at $z_3$ and anti-chiral at $z_4$, which as before is given by
\eq{
	\Xbf_a={}& \Xbf_{1(a3)}	 \,, \qquad \T_a=\T_{1(a3)} \,, \qquad \Tb_a=\Tb_{1(a3)} \,.
}
We define the cross-ratios as in \eqref{cross-ratios:good}.

A short computation shows
\eq{
	D_2^2 \frac{1}{\bk{1}{2} \bk{2}{1}} F\left( \Zbf_{1(23)}\right)={}& \frac{1}{\bk{2}{1}^{3}} {\bar \Dm}_2 F\left( \Zbf_{1(23)}\right)\,, \label{ddonf} \\
	{\bar D}_2^2 \frac{1}{\bk{1}{2} \bk{2}{1}} F\left( \Zbf_{1(23)}\right)={}& \frac{1}{\bk{1}{2}^{3}} {\Dm}_2 F\left( \Zbf_{1(23)}\right)\,, \label{ddonf2} 
}
with $\Dm$ and $\bar \Dm$ defined as in \eqref{SCCD}. The constraints at $z_3$ and $z_4$ are the same as in \eqref{chiral-constrants}. Thus, the constraints at points 2, 3 and 4 take a simple form:
\seq{}{
	D_{4\a} {\bar\Phi}(z_4)={}&0 \qquad \Rightarrow \qquad {\bar \Dm}_{4\ad} F\left( \Zbf_a\right)= 0 \,, \label{cont:1} \\	
	{\bar D}_{3\ad} \Phi(z_3)={}&0 \qquad \Rightarrow \qquad \left( {\Qm}_{2\a} + {\Qm}_{4\a} \right) F\left( \Zbf_a \right)= 0 \,, \label{cont:3} \\
	{\bar D}_{2}^2 \Jm(z_2) ={}&0 \qquad \Rightarrow \qquad {\Dm}_{2}^2 F\left( \Zbf_a\right)= 0 \,, \label{cont:4} \\	
	D_{2}^2 \Jm(z_2) ={}&0 \qquad \Rightarrow \qquad {\bar \Dm}_{2}^2 F\left( \Zbf_a \right)= 0 \,, \label{cont:2}
}
with
\eq{
	\Dm_{a\a}={}& \frac{\pd}{\pd \T_a^\a} - 2 \ri \left( \s^c \Tb_a \right)_\a \frac{\pd}{\pd \Xbf_a^c} \,, \qquad {\bar \Dm}_{a\ad} = -\frac{\pd}{\pd \Tb_a^\ad} \,, \\
	{\bar \Qm}_{a\ad} ={}& \frac{\pd}{\pd \Tb_a^\ad} + 2\ri \left( \T_a \s^c \right)_\ad \frac{\pd}{\pd \Xbf_a^c}\,, \qquad \Qm_{a\a}=-\frac{\pd}{\pd \T_a^\a} \,.
}
Note that we haven't written the constraints coming from the conservation of $\Jm(z_1)$. In order to find a simple set of equations for those constraints, we will rotate the basis for the $\Zbf$s in $F\left( \Zbf_a\right)$.

The simplest constraint to solve is \eqref{cont:1}, which implies $\Bm_i=\Cm_i=\Dm_i=\Em_i=0$ except for $\Bm_1$, $\Bm_2$, $\Bm_5$, $\Bm_6$, $\Cm_1$, $\Cm_5$, $\Cm_6$ and $\Cm_{17}$, which remain unconstrained. Chirality at $z_3$, \eqref{cont:3}, relates the $\Bm$ terms and four $\Cm$ terms,
\eq{
	{\Bm}_5={}& -{\Bm}_1 \,,{}& {\Bm}_6={}& -{\Bm}_2 \,, {}&
	{\Cm}_6={}& {\Cm}_7={\Cm}_{17} = \frac{\Xbf_2^{~2}}{\Xbf_2\cdot \Xbf_4}\Cm_1 
}
and sets to zero the remaining $\Cm$ terms. \eqref{cont:2} imposes $\Cm_1=0$, while \eqref{cont:4}, give us one differential equation
\eq{
	0={}&-\frac{1}{2} u v (u+v-1) \pd_u\pd_v {\Am}_1 -\frac{1}{2} u^2 v \pd_u^2 {\Am}_1 -\frac{1}{2} u v^2 \pd_v ^2 {\Am}_1 -\frac{1}{2} v (u-v+1) \pd_v {\Am}_1 \nn \\ {}& +\frac{1}{2} u (u-v-1) \pd_u {\Bm}_1 +\frac{1}{2} v (u-v+1)  \pd_u  {\Bm}_2  +\frac{1}{2} v (u-v+1) \pd_v  {\Bm}_1 \nn \\ {}&+ \frac{v}{2 u} \left(u (v+1)-(v-1)^2\right) \pd_v  {\Bm}_2 + {\Bm}_1 \,. \label{constraint:pde}
}

In order to apply the constraints coming from the conservation of $\Jm(z_1)$, we rewrite our coordinates and the correlator. We first note that
\eq{
	\frac{\T_{1(a3)} \Xbf_{1(b3)} \Tb_{1(c3)}}{\Xbf_{1(b3)}^2}={}& -\frac{1}{\Xbf_{3(a1)}^2 \Xbf_{3(1c)}^2} \left( \T_{3(a1)} \Xbf_{3(a1)} \Xbft_{3(b1)} \Xbf_{3(1c)} \Tb_{3(c1)} \right) \,,
}
and also
\eq{
	U={}& \frac{\Xbf_{1(23)}^2}{\Xbf_{1(43)}^2}=\frac{x_{\bar 2 3}^{~2} x_{\bar 4 1}^{~2}}{x_{\bar 2 1}^{~2} x_{\bar 4 3}^{~2}} =\frac{\Xbf_{3(41)}^2}{\Xbf_{3(21)}^2}	\,, \\ 
	V={}& \frac{\Xbf_{1(23)}\cdot\Xbf_{1(43)}}{\Xbf_{1(42)}^{~2}}=\frac{\Xbf_{3(21)}\cdot\Xbf_{3(41)}}{\Xbf_{3(21)}^{~2}}\,.
}

Thus, now we can write our correlator as,
\eq{
	\left\langle \Jm(z_1) \, \Jm(z_2) \, \Phi(z_3) \, {\bar\Phi}(z_4) \right\rangle	=&{} \frac{1}{ \bk{2}{3} \bk{3}{2} \bk{4}{3}^q \bk{1}{3} \bk{3}{1} } \left(\frac{F\left( \Zbf_{3(a1)}\right) }{ \Xbf_{3(12)}^{2} \Xbf_{3(21)}^{2} } \right) \,.
}

The constraints coming from $\Jm(z_1)$ now read
\seq{}{
	D_1^2 \Jm(z_1)=&{} 0 \qquad \Rightarrow  \qquad \left( {\bar\Qm}_2 + {\bar\Qm}_4  \right)^2  \left(\frac{F\left( \Zbf_{3(a1)}\right) }{ \Xbf_{3(12)}^{2} \Xbf_{3(21)}^{2} } \right) =0 \,, \label{cons:6}\\
	{\bar D}_1^2 \Jm(z_1) =&{} 0 \qquad \Rightarrow \qquad  \left( {\Qm}_2 + {\Qm}_4  \right)^2  \left(\frac{F\left( \Zbf_{3(a1)}\right) }{ \Xbf_{3(12)}^{2} \Xbf_{3(21)}^{2} } \right) = 0 \,. \label{cons:5}
}

It is not hard to check that \eqref{cons:5} does not give any new equation, and \eqref{cons:6} just gives us \eqref{constraint:pde} again. Therefore, finally have the four-pint function
\eq{
	\left\langle \Jm(z_1) \, \Jm(z_2) \, \Phi(z_3) \, {\bar\Phi}(z_4) \right\rangle	=&{} \frac{1}{ \bk{1}{2} \bk{2}{1} \bk{4}{3}^{q}} \left( \Am_1 + 4\ri \left [\left(222\right) - \left(422\right) \right]\Bm_1 \right. \nn \\ & \qquad   \qquad \qquad \qquad\left. + 4\ri \left [\left(242\right) - \left(442\right) \right]\Bm_2 \right) \,, \label{jjppb:4pf}
}
with \eqref{constraint:pde} relating the terms.

As seen above, the chirality constraints (\ref{cont:1},\ref{cont:3}) are very restrictive. Indeed, we can replace the currents for general long operators $\Om_1$ and $\Om_2$
\eq{
	\left\langle \Om_1 (z_1)\, \Om_2(z_2)\, \Phi(z_3) \, {\bar\Phi}(z_4)\right\rangle ={}& \frac{\left( \Xbf_{2}^{~2}\Xbf_{4}^{~2}\right)^{\frac{1}{2}({\bar q}_1 - q_2)}}{\bk{1}{2}^{q_2} \bk{2}{1}^{\bar q_2} \bk{4}{3}^{q}}  \left( \Am_1 + 4\ri \left [\left(222\right) - \left(422\right) \right]\Bm_1 \right. \nn \\
	& \qquad  \left. + 4\ri \left [\left(242\right) - \left(442\right) \right]\Bm_2 + \left[ (222)^2 \right. \right. \nn \\ & \qquad \left.\left. +2(222)(422)  +(422)^2\right] \Cm_1 \right) \label{OOppb:4pf}
}
such that the total $U(1)_R$-charge is zero, i.e., $q_1 +q_2 = {\bar q}_1 + {\bar q}_2$

It is easy to see that \eqref{jjppb:4pf} and \eqref{OOppb:4pf} corresponds to the s-channel four-point function. In order to bootstrap this system, we also need the t-channel propagator. In the Bosonic limit, this propagator is given by
\eq{
	\left\langle \Jm(z_1) \, \Phi(z_2) \, {\bar\Phi}(z_3) \, \Jm(z_4) \right\rangle|_{\t_i=\tb_i=0} ={}& \frac{1}{\bkb{1}{2}^{(q+2)/2} \bkb{3}{4}^{(q+2)/2}} \left( \frac{\bkb{2}{4}}{\bkb{1}{4}} \right)^{(2-q)/2} \nn \\ & \qquad \times \left( \frac{\bkb{1}{4}}{\bkb{1}{3}} \right)^{(q-2)/2}\Am_1(v,u) \,.
}
In order to write the full supersymmetric propagator, we first need to define a basis. In this case, we will work with $\Zbf_{4(a2)}$. In this basis, the cross-ratios which have the right bosonic limit are given by
\eq{
	u={}&\frac{U}{V} \,, & v={}&\frac{1}{V}\,, \\
	U={}&\frac{\Xbf_1^{~2}}{\Xbf_3^{~2}} \,, & V={}& 1 + \frac{\Xbf_1^{~2}}{\Xbf_3^{~2}} - 2 \frac{\Xbf_1 \cdot \Xbf_{3}}{\Xbf_3^{~2}}\,.
}

The right four-point function should have the right $\bk{1}{4}\bk{4}{1}$ factor and the right (anti-)chirality for $(z_4)z_{3}$. Thus, the full supersymmetric four point function in the t-channel is given by
\eq{
	\left\langle \Jm(z_1) \, \Phi(z_2) \, {\bar\Phi}(z_3) \, \Jm(z_4) \right\rangle ={}& \frac{1}{\bk{1}{4} \bk{4}{1} \bk{3}{2}^q } u^{-\frac{q+2}{2}} v^q\, F\left( \Zbf_a \right) \,,
}
with $F(\Zbf_a)$ given by \eqref{solf} with $2\to1$, $4\to3$. The constraints now read
\seq{}{
	D_{3\a} {\bar\Phi}(z_3)={}&0 \qquad \Rightarrow \qquad {\bar \Dm}_{3\ad} F\left( \Zbf_a\right)= 0 \,, \label{cont2:1} \\	
	{\bar D}_{2\ad} \Phi(z_2)={}&0 \qquad \Rightarrow \qquad \left( {\Qm}_{1\a} + {\Qm}_{3\a} \right) F\left( \Zbf_a \right)= 0 \,, \label{cont2:3} \\
	{\bar D}_{1}^2 \Jm(z_1) ={}&0 \qquad \Rightarrow \qquad {\Dm}_{1}^2 F\left( \Zbf_a\right)= 0 \,, \label{cont2:4} \\	
	D_{1}^2 \Jm(z_1) ={}&0 \qquad \Rightarrow \qquad {\bar \Dm}_{1}^2  \left( u^{-\frac{q+2}{2}} v^q F\left( \Zbf_a \right) \right)= 0 \,, \label{cont2:2}
}
where again, we will compute the constraints at $z_4$ after rotating the basis. Just as before, \eqref{cont2:1} set to zero all the $\Bm_i$, $\Cm_i$, $\Dm_i$ and $\Em_1$, except for $\Bm_1$, $\Bm_2$, $\Bm_5$, $\Bm_6$, $\Cm_1$, $\Cm_5$, $\Cm_6$ and $\Cm_{17}$. \eqref{cont2:3} set to zero the remaining $\Cm_i$, except for a $\Cm_6=\Cm_7=\Cm_{17}=\Xbf_1^{~2}\Cm_1/ \Xbf_1\cdot \Xbf_3$, while imposing $\Bm_5=-\Bm_1$ and $\Bm_6=-\Bm_2$. Using \eqref{cont2:4} obtain $\Cm_1=0$, while \eqref{cont2:2} impose a differential equation,
\eq{
	0={}&4 u ((q-1) (u-1)-v){\pd}_u{\Am}_1 -2 v (q (-3 u+v-1)+2 (u+v-1)) {\pd}_v{\Am}_1   \nn \\ & +4 u^2 v {\pd}_u^2{\Am}_1 +4 u v^2 {\pd}_v^2{\Am}_1 +4 u v (u+v-1) {\pd}_u {\pd}_v{\Am}_1 +4 u (u-v-1) {\pd}_u{\Bm}_1 \nn \\ & +\left(2 (q-1) ((q-2) u+q+2)-\left(q^2-4\right) v\right) {\Am}_1    +4 u (u-v-1) {\pd}_v{\Bm}_2  \nn \\ & +\frac{4 u}{v} \left((u-1)^2-(u+1) v\right)  {\pd}_u{\Bm}_2 +4 v (u-v+1)  {\pd}_v{\Bm}_1  +\frac{1}{v}\left(+4 (u+1) v \right. \nn \\ & \left.-2 q (u-1) (u-v+1)-4 (u-1)^2\right) {\Bm}_2  + (2 q (u-v+3)-4 (u-v+1)) {\Bm}_1 \,, \label{JJPPb:t_constraint}
}
Finally, imposing the constraints at $z_4$ will not give us any new constraint.

Our final four-point function is
\eq{
	\left\langle \Jm(z_1) \, \Phi(z_2) \, {\bar\Phi}(z_3) \, \Jm(z_4) \right\rangle ={}& \frac{1}{\bk{1}{4} \bk{4}{1} \bk{3}{2}^q } u^{-\frac{q+2}{2}} v^q\, \left( \Am_1 + 4\ri \left [\left(111\right) - \left(311\right) \right]\Bm_1 \right. \nn \\
	& \qquad \qquad \qquad \left. + 4\ri \left [\left(131\right) - \left(331\right) \right]\Bm_2 \right) \,, \label{jjppb:4pf_tchannel}
}
with \eqref{JJPPb:t_constraint} relating the functions of the cross-ratios.

If we replace the currents for long-multiplets, the four-point function is
\eq{
	\left\langle \Om(z_1) \, \Phi(z_2) \, {\bar\Phi}(z_3) \, \Om(z_4) \right\rangle ={}& \frac{\left(\Xbf_1^{~2}\Xbf_3^{~2}\right)^{\frac{1}{2}(\bar q_1-q_2)}}{\bk{1}{4}^{q_2} \bk{4}{1}^{\bar q_2} \bk{3}{2}^q } u^{-\frac{q+q_2+\bar q_2}{2}} v^q\, \left( \Am_1 + 4\ri \left [\left(111\right) - \left(311\right) \right]\Bm_1 \right. \nn \\
	& \qquad \qquad   + 4\ri \left [\left(131\right) - \left(331\right) \right]\Bm_2 + \left[ (111)^2  \right. \nn \\ & \qquad\qquad \left.\left. +2(111)(311)  +(311)^2\right] \Cm_1 \right) \,, \label{OOppb:4pf_tchannel}
}
with $q_1+q_2=\bar q_1+\bar q_2$.

\subsection{
\texorpdfstring{$\left\langle \Jm \, \Jm \, \Jm \, \Jm \right\rangle$}{J J J J}}

For a system solely composed of current multiplets $\Jm$, which are not necessarily the same current, we choose our basis to be $\Zbf_{1(a3)}$, with the cross-ratios given by \eqref{cross-ratios:good}. Naively, one would write the four-pint function as
\eq{
	\frac{1}{\bk{1}{2}\bk{2}{1}\bk{3}{4}\bk{4}{3}}F(\Zbf_a)\,,
}
which has the right s-channel limit. But the relations \eqref{ddonf} and \eqref{ddonf2} tell us that we should have only $\bk{i}{1}\bk{1}{i}$ terms. Thus, the supersymmetric four-point function has to be
\eq{
 	\left\langle \Jm (z_1)\, \Jm (z_1)\, \Jm (z_3) \, \Jm (z_4)\right\rangle ={}& \frac{1}{\bk{1}{2} \bk{2}{1} \bk{1}{3} \bk{3}{1} \bk{1}{4} \bk{4}{1}} \frac{F\left( \Zbf_a \right)}{\left( \Xbf_4 ^{~2}\right)^2} \,, \label{4pf:jjjj}
 }
with $F(\Zbf_a)$ given by \eqref{solf}. We can check that \eqref{4pf:jjjj} has the right conformal weights, and it also reduces in the bosonic limit to the usual s-channel four-point function.

Imposing ${D_4}^2 \Jm(z_4)=0$ implies ${\bar\Dm_4}^2 F(\Zbf_a)=0$, which implies
\eq{
	{\Cm}_2  ={\Cm}_{13}  ={\Cm}_{14}  ={\Cm}_{18}  ={\Dm}_5  ={\Dm}_6  ={\Dm}_7  ={\Dm}_8  ={\Em}_1 =0\,.
}
Imposing ${D_2}^2 \Jm(z_2)=0$ implies ${\bar\Dm_2}^2 F(\Zbf_a)=0$, which implies
\eq{
	 {\Cm}_1 =  {\Cm}_5 =  {\Cm}_6 =  {\Cm}_{17} =  {\Dm}_1 =  {\Dm}_2 =  {\Dm}_3 =  {\Dm}_4 = 0\,.
}
The conservation equation at $z_2$, ${\bar D_2}^2 \Jm(z_2)=0$ now reads $\left({\Qm_2+\Qm_4}\right)^2 F(\Zbf_a)=0$, which imposes two relations among the remaining $\Cm_i$ terms,
\eq{
	{\Cm}_3 ={}& -{\Cm}_{11} -{\Cm}_{12} -{\Cm}_{16} -\frac{u {\Cm}_8 +v {\Cm}_9}{u+v-1}\,, \\ 
	{\Cm}_{15} ={}& -{\Cm}_4 -{\Cm}_7 -{\Cm}_{10} -\frac{u {\Cm}_8 +v {\Cm}_9}{u+v-1}\,.
}

In order to impose the constraints at $z_1$, we need to rotate our basis. Following the same procedure as above, we rotate to $\Zbf_{3(a1)}$. The four-point function now reads
\eq{
	\left\langle \Jm (z_1)\, \Jm (z_1)\, \Jm (z_3) \, \Jm (z_4)\right\rangle ={}&\frac{1}{\bk{1}{3}\bk{3}{1}\bk{2}{3}\bk{3}{2}\bk{4}{3}\bk{3}{4}}\frac{\Xbf_4^{~2}}{\Xbf_2^{~2} \Xbfb_{2}^{~2} \Xbfb_{4}^{~2} } F(\Zbf_{3(a1)})\,,
}
which can be written as
\eq{
	\left\langle \Jm (z_1)\, \Jm (z_1)\, \Jm (z_3) \, \Jm (z_4)\right\rangle ={}&\frac{1}{\bk{1}{3}\bk{3}{1}\bk{2}{3}\bk{3}{2}\bk{4}{3}\bk{3}{4}} \nn \\ 
	&\qquad \qquad \times \left(1-4\ri (222) \right)\left(1-4\ri (444) \right)\frac{F(\Zbf_{3(a1)})}{(\Xbf_2^{~2})^2}\,.
}
Imposing ${D_1}^2 \Jm(z_1)=0$ implies $\left({\Qm_2+\Qm_4}\right)^2 F(\Zbf_a)=0$ in the new basis, which give us two more relations
\eq{
	(1+u-v)\Cm_{10}={}& 16 u v {\Am}_1 +16 u^2 {\Bm}_2 +16 u v {\Bm}_1 +16 u v {\Bm}_3 +16 u (v-1) {\Bm}_4 +16 u v {\Bm}_6 \nn \\ & +16 u v {\Bm}_7 +16 u v {\Bm}_8 +16 (v-1) v {\Bm}_5 +\left(-(u+1) v+u+v^2\right) {\Cm}_4 \nn \\ &+u (u-v) {\Cm}_{11} +u {\Cm}_{12} +u (u-v) {\Cm}_{16} -\frac{u (v-1) (u-v) {\Cm}_8 }{u+v-1}\nn \\ &+(u+v-1) {\Cm}_7 -\frac{(v-1) v (u-v) {\Cm}_9 }{u+v-1} \,,\\
	(1+u-v)\Cm_{11}={}& 16 u v {\Am}_1 +16 u^2 {\Bm}_4 +16 u v {\Bm}_1 +16 u (v-1) {\Bm}_2 +16 u v {\Bm}_3 +16 u v {\Bm}_5 \nn \\ &+16 u v {\Bm}_6 +16 u v {\Bm}_8 +16 (v-1) v {\Bm}_7 +u (v-u) {\Cm}_4 +u (-u+v+1) {\Cm}_7 \nn \\ &+\frac{u (v (u-v+2)-1) {\Cm}_8 }{u+v-1}+\frac{v (v (u-v+2)-1) {\Cm}_9 }{u+v-1}+(v-1) (u-v) {\Cm}_{11} \nn \\ &+(v (u-v+2)-1) {\Cm}_{12} +(v-1) (u-v) {\Cm}_{16} \,.
} 

The remaining constraint give us 16 long differential equations, see below.
Finally, the four-point function is given by
 \eq{
	\left\langle \Jm (z_1)\, \Jm (z_1)\, \Jm (z_3) \, \Jm (z_4)\right\rangle ={}& \frac{1}{\bk{1}{2} \bk{2}{1} \bk{1}{3} \bk{3}{1} \bk{1}{4} \bk{4}{1}} \frac{F_{\Jm\Jm\Jm\Jm}\left( \Zbf_a \right)}{\left( \Xbf_4 ^{~2}\right)^2} \,, \label{JJJJ:4pf}
}
with
\eq{
	F_{\Jm\Jm\Jm\Jm}\left( \Zbf_a \right)={}& \Am_1 + 4\ri \Bm_1 \left(222\right)+ 4\ri \Bm_2 \left(242\right)+ 4\ri \Bm_3 \left(224\right) + 4\ri \Bm_4 \left(244\right) + 4\ri \Bm_5 \left(422\right) \nn \\ & + 4\ri \Bm_6 \left(442\right) + 4\ri \Bm_7 \left(424\right) + 4\ri \Bm_8 \left(444\right)  + \Cm_3 \left(242\right)\left(224\right) + \Cm_4 \left(222\right)\left(244\right) \nn \\ & + \Cm_5 \left(222\right)\left(442\right) + \Cm_7 \left(244\right)\left(422\right) + \Cm_8 \left(242\right)\left(444\right)+ \Cm_9 \left(222\right)\left(424\right) \nn \\ & + \Cm_{10} \left(222\right)\left(444\right) + \Cm_{11} \left(224\right)\left(442\right) + \Cm_{12} \left(242\right)\left(424\right) + \Cm_{15} \left(422\right)\left(444\right) \nn \\ & + \Cm_{16} \left(424\right)\left(442\right) \,,
}
and
\seq{}{
	v (u-v+1) {\Cm}_{10} ={}& 16 u v {\Am}_1 +16 u^2 {\Bm}_2 +16 u v {\Bm}_1 +16 u v {\Bm}_3 +16 u (v-1) {\Bm}_4 +16 u v {\Bm}_6 \nn \\ & +16 u v {\Bm}_7 +16 u v {\Bm}_8 +16 (v-1) v {\Bm}_5 +u (v-u) {\Cm}_3 +u (v-u) {\Cm}_8 \nn \\ & -u (u-v1) {\Cm}_{12} +(v-1) (v-u) {\Cm}_4 +(u+v-1) {\Cm}_7 +v (v-u) {\Cm}_9  \,, \\
	v (u-v+1) {\Cm}_{11} ={}& 16 u v {\Am}_1 +16 u^2 {\Bm}_4 +16 u v {\Bm}_1 +16 u (v-1) {\Bm}_2 +16 u v {\Bm}_3 +16 u v {\Bm}_5 \nn \\ & +16 u v {\Bm}_6 +16 u v {\Bm}_8 +16 (v-1) v {\Bm}_7 -u (u-v) {\Cm}_4 +u (-u+v+1) {\Cm}_7 \nn \\ & +u {\Cm}_8 -(v-1) (u-v) {\Cm}_3 +v {\Cm}_9 +(u+v-1) {\Cm}_{12}  \,, \\
	v (u-v+1) {\Cm}_{15} ={}& -16 u v {\Am}_1 -16 u^2 {\Bm}_2 -16 u v {\Bm}_1 -16 u v {\Bm}_3 -16 u (v-1) {\Bm}_4 -16 u v {\Bm}_6 \nn \\ & -16 u v {\Bm}_7 -16 u v {\Bm}_8 -16 (v-1) v {\Bm}_5 +\frac{u^2 (u-v-1) {\Cm}_8 }{u+v-1}+u (u-v) {\Cm}_3 \nn \\ & -u {\Cm}_4 +\frac{u v (u-v-1) {\Cm}_9 }{u+v-1}+u (u-v-1) {\Cm}_{12} +\left((v-1)^2 \right.  \nn \\ & \left. -u (v+1)\right) {\Cm}_7  \,, \\
	v (u-v+1) {\Cm}_{16} ={}& -16 u v {\Am}_1 -16 u^2 {\Bm}_4 -16 u v {\Bm}_1 -16 u (v-1) {\Bm}_2 -16 u v {\Bm}_3 -16 u v {\Bm}_5 \nn \\ & -16 u v {\Bm}_6 -16 u v {\Bm}_8 -16 (v-1) v {\Bm}_7 -u {\Cm}_3 +u (u-v) {\Cm}_4 \nn \\ & +u (u-v-1) {\Cm}_7 +\frac{\left((v-1)^2-u (v+1)\right) }{u+v-1}\left( u  {\Cm}_8 + v  {\Cm}_9\right) \nn \\ & +\left((v-1)^2-u (v+1)\right) {\Cm}_{12}	 \,.
}

We also have several differential equations, for example
\seq{}{
	0={}& -(u+v-1) {\Cm}_3   -16 u^2 v {\pd}_v{\Bm}_4 -16 u v^2 {\pd}_u{\Bm}_4-16 u v^2 {\pd}_v{\Bm}_3 \nn \\ & -16 (v-1) v^2 {\pd}_u {\Bm}_3 -16 v {\Bm}_3  \,, \\
	0={}&-(u+v-1) {\Cm}_4  + 16 u \left(u^2-2 u-v+1\right) {\pd}_v{\Bm}_4+16 u v^2 {\pd}_u{\Bm}_3+16 (u-1) u v {\pd}_u{\Bm}_4 \nn \\ & +16 (u-1) u v {\pd}_v{\Bm}_3-16 v {\Bm}_3  \,,
}
plus several other differential equations involving derivatives of the remaining coefficients, which can be find in the attached {\texttt Mathematica} file. The equation involving the $\Cm_i$ terms are the supersymmetric generalization of the ones found in \cite{Dymarsky:2017xzb}.

\section{Casimir} \label{casimir}

Having the full supersymmetric four-point functions, we are ready to apply the super Casimir operator to them. First we write the $\Nm=1$ superconformal algebra following the conventions of \cite{Fortin:2011nq}.\footnote{Several other conventions for the $\Nm=1$ superconformal algebra can be found in the literature, see for example \cite{Park:1997bq,Bobev:2015jxa,Poland:2010wg}.} The bosonic part of the algebra is given by
\begin{subequations}\label{algebra:1}
	\begin{gather}
	\left[ M_{\m\,\n},P_{\rho} \right]= \ri \left( \eta_{\m\,\rho}P_{\n} - \eta_{\n\,\rho}P_{\m} \right)\,, \qquad \left[ M_{\m\,\n},K_{\rho} \right]= \ri \left( \eta_{\m\,\rho}K_{\n} - \eta_{\n\,\rho}K_{\m} \right)\,, \\
	\left[ M_{\m\,\n},M_{\rho\,\s} \right]= \ri \left( \eta_{\m\,\rho}M_{\n\,\s} - \eta_{\n\,\rho}M_{\m\,\s} + \eta_{\m\,\s}M_{\rho\,\n} - \eta_{\n\,\s}M_{\rho\,\m} \right) \,, \\
	\left[ D,P_{\m} \right]= -\ri P_{\m} \,, \qquad \left[ D,K_{\m} \right]= \ri K_{\m}\,, \qquad \left[ K_\m,P_{\n} \right]= 2\ri \left( \eta_{\m\,\n}D - M_{\m\,\n} \right)\,,
	\end{gather}
\end{subequations}
while the anti-commutators are
\begin{subequations}\label{algebra:2}
	\begin{gather}
	\left\lbrace Q_\a , {\bar Q}_\ad \right\rbrace = 2 \s^\m_{\a\,\ad}P_\m \,, \qquad \left\lbrace S^\a , {\bar S}^\ad \right\rbrace = 2 {\bar\s}^{\m\,\ad\,\a} K_\m \,, \\
	\left\lbrace Q_\a , S^\b \right\rbrace = - 2 \ri \left( \s^{\m\,\n} \right)_\a^{~\b}M_{\m\,\n} + \d_\a^{~\b} \left( 2\ri D+ 3 R \right) \,, \\
	\left\lbrace {\bar Q}_\ad , {\bar S}^\bd \right\rbrace = -2 \ri \left( {\bar\s}^{\m\,\n} \right)_{~\ad}^{\bd}M_{\m\,\n} - \d_{~\ad}^{\bd} \left( 2\ri D - 3 R \right) \,,
	\end{gather}
\end{subequations}
finally, the remaining of the algebra is given by
\seq{\label{algebra:3}}{
	\left[ D, Q_\a \right]={}& -\frac{\ri}{2}Q_\a \,, & \left[ D, {\bar Q}_\ad \right]={}& -\frac{\ri}{2}{\bar Q}_\ad \,, \\ 
	\left[ D, S^\a \right]={}& \frac{\ri}{2}S^\a \,, & \left[ D, {\bar S}^\ad \right]={}& \frac{\ri}{2}{\bar S}^\ad \,, \\
	\left[ R, Q_\a \right]={}& -Q_\a \,, & \left[ R, {\bar Q}_\ad \right]={}& {\bar Q}_\ad \,, \\ 
	\left[ R, S^\a \right]={}& S^\a \,, & \left[ R, {\bar S}^\ad \right]={}& - {\bar S}^\ad \,, \\
	\left[ K^\m , Q_\a \right] ={}& - \s^\m _{\a\,\ad} {\bar S}^\ad \,, & \left[ K^\m , {\bar Q}_\ad \right] ={}& \s^\m _{\a\,\ad} S^\a \,, \\
	\left[ P^\m , {\bar S}^\ad \right] ={}&  {\bar\s}^{\m\,\ad\,\a} Q_\a \,, & \left[ P^\m , {S}^\a \right] ={}&  -{\bar\s}^{\m\,\ad\,\a} {\bar Q}_\ad \,, \\
	\left[ M_{\m\,\n} , Q_\a \right] ={}& -\ri \left(\s_{\m\,\n}\right)_\a ^{~\b}Q_\b \,, & \left[ M_{\m\,\n} , {\bar Q}_\ad \right] ={}& \ri \left({\bar \s}_{\m\,\n} \right)_{~\ad}^{\bd} {\bar Q}_\bd \,, \\
	\left[ M_{\m\,\n} , S_\a \right] ={}& -\ri \left(\s_{\m\,\n}\right)_\a ^{~\b}S_\b \,, & \left[ M_{\m\,\n} , {\bar S}_\ad\right] ={}& \ri \left({\bar \s}_{\m\,\n} \right)_{~\ad}^{\bd} {\bar S}_\bd \,, 
}
and the rest of the (anti-)commutators vanish. We can represent the operators in the algebra with differential operators acting on the coordinates. Following \cite{Fortin:2011nq},
\seq{\label{diffoperators}}{
	\Pm_a = {}& - \ri \pd_a \,, \\ 
	\qquad \Qm_\a ={} & \frac{\pd}{\pd \, \t^\a} - \ri \left( \s^a \, \tb\right)_\a  \pd_a \,, \\
	{\bar\Qm}_\ad ={} & - \frac{\pd}{\pd \, \tb^\ad} +  \ri \left(\t\, \s^a \right)_\ad  \pd_a \,, \\
	\Rm = {}& \t^\a \frac{\pd}{\pd\,\t^\a} - \tb^\ad \frac{\pd}{\pd\,\tb^\ad} - \frac{2}{3} (q - {\bar q})   \,, \\
	\Dm = {}& \ri \left( x^a \pd_a + \frac{1}{2} \t^\a \frac{\pd}{\pd\,\t^\a} + \frac{1}{2} \tb^\ad \frac{\pd}{\pd\,\tb^\ad} + q + {\bar q} \right) \,, \\
	\Km_a ={} & -\ri \left( x^2\pd_a -2 x_a x^b \pd_b - 2 q x_{+a} - 2 {\bar q} x_{-a} - 2 (\t \s_b \tb)(\t \s^b \tb) \pd_a + (\t\,\s_a\,\rxt_+)^\a \frac{\pd}{\pd\,\t^\a} \right. \nn \\ & \left. + (\rxt_-\s_a\,\t)^\ad \frac{\pd}{\pd\,\tb^\ad} + 2 \ri \left( x_+^b \d^c_a - 2 \ri (\t \s_a {\bar \s}^{bc}\tb) \right)s_{bc} \right) \,, \\
	\Mm_{ab} ={} & - \ri \left( x_a \pd_b - x_b \pd_a + (\s_{ab})_\a^{~\b}\t^\a \frac{\pd}{\pd\,\t^\b} - ({\tilde\s}_{ab})_{~\ad}^{\bd}\tb^\ad \frac{\pd}{\pd\,\tb^\bd} +  \ri s_{ab}\right) \,, \\
	\Sm^\a ={}& - \left( (\t\s^a\rxt_+)^\a \pd_a +\ri \rxt_-^{\ad\a}\frac{\pd}{\pd\,\tb^\ad} + 4 \t^\a\t^\b \frac{\pd}{\pd\,\t^\b} - 4 q \t^\a -2 \t^\b s_\b ^{~\a}\right) \,, \\
	{\bar\Sm}^\ad ={}& - \left( (\rxt_- \s^a \tb)^\ad \pd_a +\ri \rxt_+^{\ad\a}\frac{\pd}{\pd\,\t^\a} + 4 \tb^\ad\tb^\bd \frac{\pd}{\pd\,\tb^\bd} - 4 {\bar q} \tb^\ad -2 \tb^\bd s_\bd ^{~\ad}\right) \,, 
}
where $q= \frac{1}{2}\left(\D+\frac{3}{2}r \right)$, ${\bar q}= \frac{1}{2}\left(\D-\frac{3}{2}r \right)$, with $\D$ being the scaling dimension and $r$ the $U(1)_R$-charge of the operator where the generators act upon, $s_{ab}$ is the proper rotation matrix and $s_{\a\b}$ and $s_{\ad\bd}$ are the proper projections.

The quadratic super Casimir for this algebra is
\eq{
	C^{(2)} ={}& \frac{1}{2} M^{a\,b}M_{a\,b}-D^2 - \frac{1}{2}P^a\, K_a - \frac{1}{2} K^a\, P_a -\frac{3}{4}R^2 - \frac{1}{4} S^\a \, Q_\a + \frac{1}{4} Q_\a \, S^\a \nn \\ & + \frac{1}{4} {\bar S}^\ad \, {\bar Q}_\ad - \frac{1}{4} {\bar Q}_\ad \, {\bar S}^\ad, \label{Cas2}
}
which for our generators reads
\eq{
	C^{(2)} ={}& \frac{1}{2} \Mm^{a\,b}\Mm_{a\,b}-\Dm^2 - \frac{1}{2}\Pm^a\, \Km_a - \frac{1}{2} \Km^a\, \Pm_a -\frac{3}{4}\Rm^2 + \frac{1}{4} \Sm^\a \, \Qm_\a - \frac{1}{4} \Qm_\a \, \Sm^\a \nn \\ & - \frac{1}{4} {\bar \Sm}^\ad \, {\bar \Qm}_\ad + \frac{1}{4} {\bar \Qm}_\ad \, {\bar \Sm}^\ad. \label{Cas2:operator}
}

The eigenvalue of a supermultiplet under \eqref{Cas2:operator} is
\eq{
	\lambda= \frac{1}{2}j(j+2) +\frac{1}{2}{\bar j}({\bar j}+2)+(q+{\bar q})(q+{\bar q}-2)-\frac{1}{3}(q-{\bar q})^2 \,.\label{sCas:eigenvalue}
}

When acting on two different points, \eqref{Cas2:operator} acts as a differential operator (see \eqref{diffoperators}.) Explicitly, when acting on two scalar supermultiplets, \eqref{Cas2:operator} reads 
\eq{
	C^{(2)}_{ij} ={}& \frac{\ri}{2}\rxt^{\ad\a}_{{\bar i}j}D_{j\a}{\bar D}_{i\ad} + \frac{\ri}{2}\rxt^{\ad\a}_{{\bar j}i}D_{i\a}{\bar D}_{j\ad} - 2 \t_{ij}^\a \t_{ij}^\b D_{j\a}D_{i\b} - 2 \tb_{ij}^\ad \tb_{ij}^\bd {\bar D}_{j\ad}{\bar D}_{i\bd} + 2 x_{j\bar i}^a \t_{ij}^\a \pd_{i a}D_{j\a} \nn \\ 
	& - 2 x_{i\bar j}^a \t_{ij}^\a \pd_{j a}D_{i\a} + 2 x_{i\bar j}^a \tb_{ij}^\ad \pd_{i a}{\bar D}_{j\ad} - 2 x_{j\bar i}^a \tb_{ij}^\ad \pd_{j a}{\bar D}_{i\ad} + \left( {\bar \s}^a \rx_{i\bar j}\tb_{ij}\right)^\ad \pd_{ia}{\bar D}_{j\ad} \nn \\ 
	& - \left( {\bar \s}^a \rx_{j\bar i}\tb_{ij}\right)^\ad \pd_{ja}{\bar D}_{i\ad} - \left( \t_{ij} \rx_{i\bar j}\s^a \right)^\a \pd_{j a}D_{i\a} + \left( \t_{ij} \rx_{j\bar i}\s^a \right)^\a \pd_{i a}D_{j\a} \nn \\ 
	& + 2 \ri \left(\t_{ij}\s^a \rxt_{\bar i j} \s^b \tb_{ij} \right) \pd_{ia}\pd_{jb} -2 x_{j\bar i}^a x_{j\bar i}^b\pd_{ia}\pd_{jb} + \la {\bar i} j \ra \eta^{ab}\pd_{ia}\pd_{jb} + \frac{1}{2}\left( \d_{ij} -4q_i \right)x_{i\bar j}^a \pd_{ja} \nn \\ 
	& + \frac{1}{2}\left( \d_{ij} -4q_j \right)x_{j\bar i}^a \pd_{ia} + 2 \left( \d_{ij} - q_i \right) \t_{ij}^\a D_{j\a} -  2 \left( \d_{ij} - q_j \right) \t_{ij}^\a D_{i\a} + 2 \left( \d_{ij} - {\bar q}_j \right) \tb_{ij}^\ad {\bar D}_{i\a} \nn \\ 
	& - 2 \left( \d_{ij} - {\bar q}_i \right) \tb_{ij}^\ad {\bar D}_{j\a} + \frac{1}{2} \left( 4 {\bar q}_i  - 3 \d_{ij} \right) x_{j\bar i}^a \pd_{ja} + \frac{1}{2} \left( 4 {\bar q}_j  - 3 \d_{ij} \right) x_{i\bar j}^a \pd_{ia}  \nn \\ 
	& + \frac{2}{3} \left((q_i+q_j)^2 + ({\bar q}_i + {\bar q}_j)^2 - 3 ({\bar q}_i+{\bar q}_j) + (q_i + q_j)(4 {\bar q}_i + 4 {\bar q}_j -3 )\right) \nn \\
	& - 2 \d_{ij} (q_i +q_j +{\bar q}_i +{\bar q}_j ) \,. \label{sCas}
}

Although it would be nice to write \eqref{sCas} in term of derivatives acting on $\Zbf$ (see \eqref{SCCD}) we found that this is not as systematic nor straightforward as one might expect. Instead, it is simpler to use the superconformal transformations and send one coordinate to $z_j=(\infty,0,0)$. In this frame, the pair of $\Zbf_a$ coordinates is easily written as a pair of $z_{ij}$ coordinates.

\subsection{\texorpdfstring{$\left\langle \Phi \, {\bar\Phi} \, \Phi \, {\bar\Phi} \right\rangle$}{phi phidagger phi phidagger}}

As mentioned above, out aim is to study the simplest four-point functions with non-zero nilpotent invariants, which are \eqref{jjppb:4pf}, \eqref{OOppb:4pf}, \eqref{jjppb:4pf_tchannel} and \eqref{OOppb:4pf_tchannel}. Before we proceed, it is instructive to study the $\left\langle \Phi \,\bar \Phi \,\Phi \,\bar\Phi \right\rangle$ propagator. We already know that the propagator is given by
\eq{
	\left\langle \Phi(z_1)\,\bar \Phi(z_2)\,\Phi(z_3)\,\bar\Phi(z_4) \right\rangle ={}& \frac{1}{\bk{2}{1}^{q} \bk{4}{3}^{q}} \Am_1(u,v)\,. \tag{\ref{4pf:pbppbp_sol}}
}
Expanding $\Am_1$ as a sum of superconformal blocks, now it reads
\eq{
	\Am_1 ={}& \sum_\Om |\l_{\Phi {\bar\Phi}\Om}|^2\, \Gm_{0,\Om}^{0}(u,v)\,. \label{scpw:chirals}
}

When acting with the quadratic super Casimir \eqref{sCas} on it, we find the eigenvalue equation
\eq{
	\l\, \Gm_{0,\Om}^{0} ={}& \left(-2 (2 u v+u+v (3-2 v)-1) \pd_v +2 v \left((v-1)^2-u (v+1)\right)\pd_v^2  \right. \nn \\
	&\left. -4 u (1 + u - v) \pd_u -4 u v (u-v+1) \pd_u\pd_v +2 u^2 (-u+v+1) \pd_u^2  \right)\, \Gm_{0,\Om}^{0} \,, \label{sCas:chirals}
}
with $\l$ given in \eqref{sCas:eigenvalue}. This equation was already found in \cite{Fitzpatrick:2014oza} using the super-embedding formalism.\footnote{In fact, in \cite{Fitzpatrick:2014oza} it was obtained the super Casimir equation for general $\Nm$ in a compact form.} In order to solve this equation, we look at the three-point function \cite{Poland:2010wg}. The only solution for the three-point function consistent with the constraints is
\eq{
	\left \langle \Phi(z_1) \,\bar \Phi(z_2) \, \Om(z_3)^{(\ell)} \right\rangle ={}& \frac{\l_{\Phi\bar\Phi\Om}}{\bk{3}{1}^q\bk{2}{3}^q}  \frac{\Xbf_{3(21)}^{\ell}}{\left(\Xbf_{3(21)}^{~2}\right)^{(2q+\ell-\D)/2}} \,, \label{PPb:3pf}
}
where $\Xbf^{\ell}$ denotes symmetric-traceless and $\D$ is the dimension of $\Om$. Looking at \eqref{4pf:pbppbp_sol}, we see that $\Am_1$ is proportional to the $\left\langle \phi \phi^* \phi \phi^* \right\rangle$ four-point function. From \eqref{PPb:3pf} we can read which components of the $\Om^{(\ell)}$ supermultiplet appear in the $\phi\phi^*$ OPE. In order to do so, we set $\t_{1,2}=\tb_{1,2}=0$ and expand around the $\t_3$ and $\tb_3$ components in \eqref{PPb:3pf}:
\eq{
	\left.  \left \langle \Phi(z_1) \,\bar \Phi(z_2) \, \Om(z_3)^{(\ell)} \right\rangle \right|_{\t_{1,2}=\tb_{1,2}=0}={}&  \left \langle \phi(x_1) \,\phi^*(x_2) \, \Om(z_3)^{(\ell)} \right\rangle \nn \\
	={}&\frac{\l_{\Phi\bar\Phi\Om}}{(\bkb{3}{1}\bkb{3}{2})^q} \left(\frac{Z^{\ell}}{\left(Z^2\right)^{(2q+\ell-\D)/2}} + \cdots \right) \,, \label{PPb:3pf_expansion}
}
where
\eq{
	Z^a=\frac{x_{31}^a}{\bkb{3}{1}}-\frac{x_{32}^a}{\bkb{3}{2}}\,,
}
and the $\cdots$ terms in \eqref{PPb:3pf_expansion} correspond to terms proportional to $(\t_3\tb_3)$ and $(\t_3\tb_3)^2$. By looking at \eqref{PPb:3pf_expansion}, we see that the $\t_3=\tb_3=0$ term correspond to an operator with $(\ell,\ell)$ spin and conformal dimension $\D$, and the $(\t_3\tb_3)$ and $(\t_3\tb_3)^2$ terms correspond to $\Qm{\bar\Qm}$ and $(\Qm{\bar\Qm})^2$ descendants of this operator, respectively. In general, for an operator with ${(j,\bar j)}$ spin and conformal dimension $\Delta$, will have four $\Qm{\bar\Qm}$ descendants with spin $(j + 1,\bar j+1)$ , $(j +1,\bar j-1)$, $(j-1,\bar j+1)$ and $(j-1,\bar j-1)$, all of them with conformal dimension $\Delta+1$, and it will only have one $(\Qm{\bar\Qm})^2$ descendant with the same $(j,\bar j)$ spin and conformal dimension $\D+2$. Thus, the $\phi\times \phi^*$ OPE is\footnote{Note that in any CFT, two scalars can only exchange traceless symmetric tensors, this is why only four operators appear in this OPE instead of six.}
\eq{
	\phi \times \phi^* \sim A^{(\ell)}_{\D} + B^{(\ell+1)}_{\D+1} + C^{(\ell-1)}_{\D+1} + D^{(\ell)}_{\D+2}\,. \label{OPE:ppb}
}

With this information, we can make an ansatz for $\Gm_{0,\Om}^{0}$:
\eq{
	\Gm_{0,\Om}^{0} (u,v) = g_{\D,\ell}(u,v) + c_1 g_{\D+1,\ell+1}(u,v) + c_2 g_{\D+1,\ell-1}(u,v) + c_3 g_{\D+2,\ell}(u,v), \label{PPbPPB:ansatz}
}
where $g_{\D,\ell}(u,v)$ are the conformal blocks \cite{Dolan:2000ut,Dolan:2003hv}
\eq{
	g_{\a,\b}(u,v)={}& g_{\a,\b}^{0,0}(u,v) \,,
}
and
\eq{
	g_{\a,\b}^{\gamma,\rho}(u,v) ={}& (-1)^\b \frac{z\bar z}{z-\bar z} \left(k_{\a+\b}^{\gamma,\rho}(z)k_{\a-\b-2}^{\gamma,\rho}(\bar z) - k_{\a+\b}^{\gamma,\rho}(\bar z)k_{\a-\b-2}^{\gamma,\rho}(z) \right) \,, \\
	k_\a ^{\gamma,\rho}(x) ={}& x^{\a/2}{}_2 F_1 \left(\frac{1}{2}(\a-\gamma),\frac{1}{2}(\a-\rho),\a;x\right) \,,
}
with $u=z\bar z$ and $v=(1-z)(1-\bar z)$.

Plugging this ansatz into \eqref{sCas:chirals}, we can fix the $c_i$ coefficients:
\seq{\label{PPb:coeffs}}{
	c_1 ={}& -\frac{\D+\ell}{2(\D+\ell+1)}\,, \\
	c_2 ={}& -\frac{\D-\ell-2}{8(\D-\ell-1)}\,, \\
	c_3 ={}& c_1 c_2\,,
}
which agrees with the results found originally in \cite{Poland:2010wg}. We stress that in \eqref{PPbPPB:ansatz}, the cross-ratios are the supersymmetric cross ratios and not just the bosonic limit.

\subsection{\texorpdfstring{$\left\langle \Jm \, \Jm \, \Phi \, {\bar\Phi} \right\rangle$}{J J phi phidagger}}
\subsubsection*{s-channel}
More interesting is the case of mixed operators. We begin studying
\eq{
	\left\langle \Jm(z_1) \, \Jm(z_2) \, \Phi(z_3) \, {\bar\Phi}(z_4) \right\rangle	=&{} \frac{1}{ \bk{1}{2} \bk{2}{1} \bk{4}{3}^{q}} \left( \Am_1 + 4\ri \left [\left(222\right) - \left(422\right) \right]\Bm_1 \right. \nn \\ & \qquad   \qquad \qquad \qquad\left. + 4\ri \left [\left(242\right) - \left(442\right) \right]\Bm_2 \right) \,. \tag{\ref{jjppb:4pf}}
}
As mentioned before, this propagator corresponds to the supersymmetric version of the s-channel.

In order to solve the super Casimir equations, we first expand our four-point functions as a sum of superconformal blocks 
\seq{}{
	\Am_1 ={}& \sum_\Om \Gm_{0,\Om}^{1}(u,v) \,, \\
	\Bm_1 ={}& \sum_\Om \Gm_{1,\Om}^{1}(u,v) \,, \\
	\Bm_2 ={}& \sum_\Om \Gm_{2,\Om}^{1}(u,v) \,.
}
Applying the super Casimir \eqref{sCas},\footnote{As we mentioned above, the simplest way to obtain the super Casimir equations is to send one coordinate to $(\infty,0,0)$. In this case, we choose to send $z_2$ to this particular point and we act with the super Casimir at $z_3$ and $z_4$.} we find three eigenvalue equations:
\seq{\label{JJPPb:CasEq}}{
	\l \,\Gm_{0,\Om}^{1} = {}& 2 u^2 (-u+v+1) {\pd}_u^2  \Gm_{0,\Om}^{1} -2 \left(2 u v+u-2 v^2+3 v-1\right) {\pd}_v  \Gm_{0,\Om}^{1} \nn \\ &-4 u (u-v+1)  {\pd}_u  \Gm_{0,\Om}^{1}  -4 u v (u-v+1) {\pd}_u {\pd}_v  \Gm_{0,\Om}^{1}+2 v \left((v-1)^2 \right. \nn \\ & \left.-u (v+1)\right) {\pd}_v^2  \Gm_{0,\Om}^{1} +\frac{2 (u+v-1) \Gm_{1,\Om}^{1} }{v} +4 \Gm_{2,\Om}^{1} \,,     \\
	\l \,\Gm_{1,\Om}^{1} = {}& 2 u^2 (-u+v+1) {\pd}_u^2  \Gm_{1,\Om}^{1} -4 u^2 {\pd}_u  \Gm_{1,\Om}^{1} -4 u v {\pd}_u  \Gm_{2,\Om}^{1} -4 u v (u-v+1) {\pd}_u {\pd}_v  \Gm_{1,\Om}^{1} \nn \\ & +2 (-2 u v+u+v-1) {\pd}_v  \Gm_{1,\Om}^{1} +2 v \left((v-1)^2-u (v+1)\right) {\pd}_v^2  \Gm_{1,\Om}^{1} -4 v^2 {\pd}_v  \Gm_{2,\Om}^{1} \nn \\ & -\frac{2 (u+v-1) \Gm_{1,\Om}^{1} }{v} \,,     \\
	\l \,\Gm_{2,\Om}^{1} = {}& 2 u^2 (-u+v+1) {\pd}_u^2  \Gm_{1,\Om}^{1} -4 u^2 {\pd}_u  \Gm_{1,\Om}^{1} -4 u v {\pd}_u  \Gm_{2,\Om}^{1} -4 u v (u-v+1) {\pd}_u {\pd}_v  \Gm_{1,\Om}^{1} \nn \\ & +2 (-2 u v+u+v-1) {\pd}_v  \Gm_{1,\Om}^{1} +2 v \left((v-1)^2-u (v+1)\right) {\pd}_v^2  \Gm_{1,\Om}^{1} -4 v^2 {\pd}_v  \Gm_{2,\Om}^{1} \nn \\ & -\frac{2 (u+v-1) \Gm_{1,\Om}^{1} }{v} \, .
}
Before we proceed, we point out that Casimir equations where different structures mix were first found when studying seed blocks \cite{Echeverri:2016dun}. Unlike them, we do not find a nice ``nearest-neighbor interaction'' interpretation. The only differential operators in the $\Gm_{0,\Om}^{1}$ eigenequation acts on $\Gm_{0,\Om}^{1}$ itself, while there is no $\Gm_{0,\Om}^{1}$ term in the $\Gm_{1,\Om}^{1}$ and $\Gm_{2,\Om}^{1}$ eigenvalue equations. The absence of the $\Gm_{0,\Om}^{1}$ function in the $\Gm_{1,\Om}^{1}$ and $\Gm_{2,\Om}^{1}$ eigenequations is particular to this channel, as we will see later.

Just as we did in the previous section, in order to solve these equations, we will look at the three-point functions of various descendants. The superfields $\Jm$, $\Phi$ and an external long multiplet $\Om_{(p,\bar p)}$ can be expanded as
\seq{\label{multiplets_expansion}}{
	\Jm={}& J + e_1\,\t j + e_2 \,\tb {\bar j} + e_3 \left( \t \s^\m \tb \right)j_\m \cdots \,,\\
	\Phi={}& \phi + e_4\,\t \psi +e_5 \t^2 F+\cdots \,, \\
	\Phi^\dagger={}& \phi^* + e_6\,\tb \bar\psi +e_7 \tb^2 F^* +\cdots \,, \\
	\Om_{(p,{\bar p})} ={}& m_{p+{\bar p}} + e_8 \, \t \,n_{(p+\bar p+1/2)} + e_9\, \tb\,r_{(p+\bar p+1/2)} + e_{10} \t^2 o_{p+\bar p+1} +e_{11} \tb^2 o^*_{p+\bar p+1} +\cdots \,,
}
with $e_i$ being arbitrary constants. Since there are several different normalizations for the supersymmetric descendants in the literature, we will not fix the $e_i$ constants.\footnote{For a nice discussion regarding the normalization of the two-point function of the supersymmetric descendants in $\Nm=1$ SCFTs, see \cite{Li:2014gpa}.} 

The contribution for the $\Gm_{1,\Om}^{1}$ and $\Gm_{2,\Om}^{1}$ superblocks can be read from the $\la J j_\a \phi \bar\psi_\ad \ra$ correlator:\footnote{We can also read the contribution from other correlators, any choice will be related by supersymmetry. For example, one could also use
\eq{
	\la {\bar j}_\ad J \psi_\a \phi^*\ra ={}& 4 a_0 \left[\left( \frac{1}{u^{1/2}}\Bm_2 + \frac{u^{1/2}}{v} \Bm_4 \right){\mathbb T}^4_{\a\ad} + \frac{u^{1/2}}{v} \Bm_4 {\mathbb T}^2_{\a\ad} \right] \,, \label{4pf2}
}
with
\eq{
	{\mathbb T}^3 \equiv \Km_5 {\mathbb I^{13}} \,, \qquad {\mathbb T}^4 \equiv \Km_5 {\mathbb I^{13}_{24}} \,, \qquad \Km_5= \frac{\la24\ra^{1/2}}{\left(\la13\ra \la 12 \ra^5 \la 34\ra^{2q+1}\right)^{1/2}} \,.
}
} 
\eq{
	\la J j_\a \phi \bar\psi_\ad \ra ={}& 4 a_0\left[\left( \frac{1}{u^{1/2}}\Bm_2 + \frac{u^{1/2}}{v} \Bm_4 \right){\mathbb T}^2_{\a\ad} + \frac{u^{1/2}}{v} \Bm_4 {\mathbb T}^1_{\a\ad} \right] \,, \label{4pf1}
}
where \cite{Cuomo:2017wme}
\eq{
	{\mathbb T}^1 \equiv \Km_4 {\mathbb I^{42}} \,, \qquad {\mathbb T}^2 \equiv \Km_4 {\mathbb I^{42}_{31}}  \,, \qquad \Km_4= \frac{\la13\ra^{1/2}}{\left(\la24\ra \la 12 \ra^5 \la 34\ra^{2q+1}\right)^{1/2}} \,,
}
and $a_0=\frac{\ri}{e_1 \,e_6}$. As we will see later, any non-zero $a_0$ normalization can be chosen for the superblocks. 

Since the only operator that can be exchanged is an $\Om^{(\ell)}_\D$, see \eqref{PPb:3pf}, the only three-point function for the current multiplet that we need to study is \cite{Fortin:2011nq,Berkooz:2014yda,Khandker:2014mpa}
\eq{
	\left \langle \Jm (z_1) \, \Jm (z_2) \, \Om^{(\ell)}_\D (z_3) \right \rangle ={}& \frac{1}{\bk{3}{1}\bk{1}{3}\bk{2}{3}\bk{3}{2}} \left[ \l_{\Jm\Jm\Om}^{(+)} t_+^{(\ell)} + \l_{\Jm\Jm\Om}^{(-)} t_-^{(\ell)} \right]\, \label{3pf:JJO}
}
with
\seq{}{
	t_+^{(\ell)} ={}& \frac{\Xbf_+ ^{(\ell)}}{\left(\Xbf\cdot \Xbfb \right)^{2-(\D-\ell)/2}}\left(1 - \frac{1}{4} (\D-\ell-4)(\D+\ell-6) \frac{\T^2\Tb^2}{\Xbf\cdot \Xbfb} \right) \,, \\
	t_-^{(\ell)} ={}& \frac{\Xbf_+ ^{(\ell-1)}}{\left(\Xbf\cdot \Xbfb \right)^{2-(\D-\ell)/2}}\left(\Xbf_-^{(1)} - \frac{(\D-\ell-4)}{\D-2} \frac{\Xbf_+\cdot \Xbf_-}{\Xbf\cdot \Xbfb}\Xbf_+^{(1)} \right) \,, 
}
and
\eq{
	\Xbf_+={}\frac{1}{2} \left( \Xbf_{3(12)}+\Xbfb_{3(12)}\right)\,, \qquad \Xbf_-={} \ri \left( \Xbf_{3(12)}-\Xbfb_{3(12)}\right)\,.
}
Note that under the exchange $z_1 \leftrightarrow z_2$, $t_\pm^{(\ell)}\to \pm(-)^\ell t_\pm^{(\ell)}$. For simplicity, we will only consider the case where the current multiplets are the same. Thus, for even spin, only $t_+$ contributes, while for odd spin only $t_-$ contributes.

Following a similar procedure as the one discussed below \eqref{PPb:3pf}, we know that in order to find the $J \times J$ OPE, we need to set $\t_{1,2}=\tb_{1,2}=0$ in the three-point function \eqref{3pf:JJO}. Just as in \eqref{PPb:3pf_expansion}, we find that the first term in this expansion corresponds to an operator with $(\ell,\ell)$-spin and conformal dimension $\D$ which is the lowest component of the $\Om^{(\ell)}_\D$ supermultiplet, and this term is proportional to $\l_{\Jm\Jm\Om}^{(+)}$. After setting $\t_{1,2}=\tb_{1,2}=0$, the $\T_{3(12)}\Tb_{3(12)}$ term is now proportional to $\t_3\tb_3$, thus we find that the first term proportional to the $\l_{\Jm\Jm\Om}^{(-)}$ three-point function coefficient are the $\Qm{\bar \Qm}$ descendants of the lowest component of $\Om^{(\ell)}_\D$. Therefore, the OPEs between the lowest components of $\Jm$ are
\seq{\label{OPE:jj}}{
	J \times J \sim{}& A^{(\ell)}_\D + D^{(\ell)}_{\D+2}\,,  \qquad &&\text{$\ell$ even} \,, \\
	J \times J \sim{}& B^{(\ell+1)}_{\D+1} + C^{(\ell-1)}_{\D+1}\,,  \qquad && \text{$\ell$ odd} \,,
}
where we have assume that the current multiplets are the same, as mentioned above.

Given the \eqref{OPE:ppb} and \eqref{OPE:jj} OPEs, we can make an ansatz for $\Am_1$:
\seq{\label{JJ:ansatz}}{
	\Gm_{0,\Om}^{1} ={}& g_{\D,\ell} + c_1 g_{\D+2,\ell}\,, && \text{$\ell$ even} \,, \label{JJ:ansat_even} \\
	\Gm_{0,\Om}^{1} ={}& g_{\D+1,\ell+1} + c_2 g_{\D+1,\ell-1}\,, && \text{$\ell$ odd} \,. \label{JJ:ansat_odd}
}

What about the $\Gm_{1,\Om}^{1}$ and $\Gm_{2,\Om}^{1}$ functions? From \eqref{4pf1} we know that the first $\Qm$ descendant of $\Jm$ and $\Phi$ will give us an ansatz for them. Thus, in the case of the $\psi_\a \times \phi^*$ OPE, instead of setting $\t_{1,2}=\tb_{1,2}=0$ in the three point function \eqref{PPb:3pf}, we look set $\t_2=\tb_{1,2}=0$ and study the term proportional to $\t_1$. From \eqref{PPb:3pf}, we find that the non-zero terms proportional to $\t_1$ are proportional to $\tb_3$ and $\tb_3^2 \t_3$. This implies, that in the three-point function \eqref{PPb:3pf} only $\bar \Qm$ and $\bar \Qm^2 \Qm$ descendants of the lowest component of the long multiplet will appear in the $\psi_\a \times \phi^* $ OPE. Thus, the relevant OPE is\footnote{Another way to arrive to this conclusion is to note that \eqref{PPb:3pf} has vanishing $U(1)_R$-charge, therefore, only operators with $-1$ $U(1)_R$-charge in the long multiplet can appear in the  $\psi_\a \times \phi^* $ OPE. These terms corresponds to the ones listed in \eqref{OPE:psiphi}.}
\eq{
	\psi_\a \times \phi^* \sim {}& \xi_{\D+1/2}^{(\ell+1,\ell)} + \xi_{\D+1/2}^{(\ell-1,\ell)} + \chi_{\D+3/2}^{(\ell,\ell+1)} + \chi_{\D+3/2}^{(\ell,\ell-1)} \,. \label{OPE:psiphi}
}
A similar analysis to \eqref{PPb:3pf} gives us the OPE between the lowest component of $\Jm$ and its $\Qm$-descendant:
\eq{
	J \times j_\a \sim{} & \xi_{\D+1/2}^{(\ell+1,\ell)} + \xi_{\D+1/2}^{(\ell-1,\ell)} + \chi_{\D+3/2}^{(\ell,\ell+1)} + \chi_{\D+3/2}^{(\ell,\ell-1)} \,. \label{OPE:Jj}
}
This is the same expression for both even and odd spin.

The exchange of an operator in the $\{\Delta,(\ell+p,\ell) \}$ representation of the conformal group will generate a $W^{seed}_{\Delta,\ell,p}$ seed block, while an operator in the $\{\Delta,(\ell,\ell+p) \}$ representation, which has the same Casimir eigenvalue, will generate the corresponding $W^{dual \, seed}_{\Delta,\ell,p}$ dual seed block \cite{Echeverri:2016dun} (see also \cite{Cuomo:2017wme}). Thus, the OPEs \eqref{OPE:psiphi} and \eqref{OPE:Jj} together with \eqref{4pf1} tell us\footnote{We implement the seed blocks using the {\texttt Mathematica} package provide by \cite{Cuomo:2017wme}, which can be found in \href{https://gitlab.com/bootstrapcollaboration/CFTs4D}{https://gitlab.com/bootstrapcollaboration/CFTs4D}.}
\eq{
	4a_0\left[\left( \frac{1}{u^{1/2}}\Gm_{2,\Om}^{1} + \frac{u^{1/2}}{v} \Gm_{1,\Om}^{1} \right){\mathbb T}^4_{\a\ad} + \frac{u^{1/2}}{v} \Gm_{1,\Om}^{1} {\mathbb T}^2_{\a\ad} \right] ={}& c_3 W^{0,seed}_{\D+1/2,\ell,1} + c_4 W^{0,dual\,seed}_{\D+1/2,\ell-1,1} \nn \\ & + c_5 W^{0,seed}_{\D+3/2,\ell-1,1} +c_6 W^{0,dual\,seed}_{\D+3/2,\ell,1} \,, \label{Bs:ansats_JJPPB}
}
with
\seq{}{
	W^{0,seed}_{\D,\ell,1}={}& W^{seed}_{\langle Jj_\a \Om_\D^{(\ell,\ell+1)} \rangle \langle \Om_\D^{(\ell+1,\ell)} \phi \bar\psi_\ad \rangle} \,, \\
	W^{0,dual\,seed}_{\D,\ell,1}={}& W^{seed}_{\langle Jj_\a \Om_\D^{(\ell+1,\ell)} \rangle \langle \Om_\D^{(\ell,\ell+1)} \phi \bar\psi_\ad \rangle} \,.
}

Plugging \eqref{JJ:ansat_even} and \eqref{Bs:ansats_JJPPB} in \eqref{JJPPb:CasEq}, we find
\seq{\label{JJPPb:even}}{
	c_1={}& -\frac{(\D-2) (\D-\ell -2) (\D+\ell )}{16 \D (\D-\ell -1) (\D+\ell +1)} \,, \\
	c_3={}& \ri a_0 (\D+\ell ) \,, \\
	c_4={}& \frac{\ri a_0 \ell  (-\D+\ell +2)}{\ell +1} \,, \\
	c_5={}& -\frac{\ri a_0 (\D-2) \ell  (\D-\ell -2) (\D+\ell )}{4 (\D-1) (\ell +1) (\D-\ell -1)} \,, \\
	c_6={}& \frac{\ri a_0 (\D-2) (\D-\ell -2) (\D+\ell )}{4 (\D-1) (\D+\ell +1)} \,.
}
If we plug the ansatz for odd spin, \eqref{JJ:ansat_odd}, instead of the even spin ansatz, we find
\seq{\label{JJPPb:odd}}{
	c_2={}& -\frac{ (\ell +2) (\D-\ell -2) (\D+\ell +1)}{\ell  (\D-\ell -1) (\D+\ell )} \,, \\
	c_3={}& 4 \ri a_0  (\D+\ell +1) \,, \\
	c_4={}& -\frac{4 \ri a_0  (\ell +2) (-\D+\ell +2) (\D+\ell +1)}{(\ell +1) (\D+\ell )} \,, \\
	c_5={}& -\frac{\ri a_0  \D (\ell +2) (\D-\ell -2) (\D+\ell +1)}{(\D-1) (\ell +1) (\D-\ell -1)} \,, \\
	c_6={}& -\frac{\ri a_0  \D (\D-\ell -2)}{\D-1} \,.
}
The $c_1$ and $c_2$ terms agree with \cite{Li:2017ddj}. 

A few comments are in order before we proceed. First, we set the norm of the first block always as one. The super Casimir equations are independent of such normalization. In order to properly fix it, one can, for example, use the techniques from \cite{Khandker:2014mpa}. Second, in order to find either \eqref{JJPPb:even} or \eqref{JJPPb:odd}, we only needed to solve the eigen equations for $\Gm_{0,\Om}^{1}$ and $\Gm_{1,\Om}^{1}$. The equation for $\Gm_{2,\Om}^{1}$ was automatically satisfied.\footnote{We could have also used $\Gm_{2,\Om}^{1}$ instead of $\Gm_{1,\Om}^{1}$.} Third, as we mentioned before, the blocks coming from the lowest component of the multiplets are independent of the $a_0$ normalization.

\subsubsection*{t-channel}

Now we turn our attention to the t-channel superblocks. We already computed the four-point function in the previous section,
\eq{
	\left\langle \Jm(z_1) \, \Phi(z_2) \, {\bar\Phi}(z_3) \, \Jm(z_4) \right\rangle ={}& \frac{1}{\bk{1}{4} \bk{4}{1} \bk{3}{2}^q } u^{-\frac{q+2}{2}} v^q\, \left( \Am_1 + 4\ri \left [\left(111\right) - \left(311\right) \right]\Bm_1 \right. \nn \\
	& \qquad \qquad \qquad \left. + 4\ri \left [\left(131\right) - \left(331\right) \right]\Bm_2 \right) \,. \tag{\ref{jjppb:4pf_tchannel}}
}
We expand this correlator as a sum of blocks
\seq{}{
	\Am_1 ={}& \sum_\Om \Gm_{0,\Om}^{2}(u,v) \,, \\
	\Bm_1 ={}& \sum_\Om \Gm_{1,\Om}^{2}(u,v) \,, \\
	\Bm_2 ={}& \sum_\Om \Gm_{2,\Om}^{2}(u,v) \,.
}

As before, we apply \eqref{sCas} to \eqref{jjppb:4pf_tchannel}, and we find three eigen equations
\seq{\label{JJPPb:t_channel}}{
	\l\, \Gm_{0,\Om}^{2} ={}& +2 \left(q \left((v-1)^2-u (v+1)\right)+u+v-1\right) {\pd}_v\Gm_{0,\Om}^{2} -4 u v (u-v+1) {\pd}_u {\pd}_v\Gm_{0,\Om}^{2} \nn \\ 
		& -\frac{2 (u+v-1) \Gm_{2,\Om}^{2}}{v} -2 q u (u-v+1) {\pd}_u\Gm_{0,\Om}^{2} +2 u^2 (-u+v+1) {\pd}_u^2\Gm_{0,\Om}^{2} \nn \\ 
		& +\frac{1}{6} q (-3 q u+3 (q-2) v-5 q+6 u+6) \Gm_{0,\Om}^{2} -4 \Gm_{1,\Om}^{2}+2 v \left((v-1)^2 \right. \nn \\
		& \left. -u (v+1)\right) {\pd}_v^2\Gm_{0,\Om}^{2}   \\
	\l\, \Gm_{1,\Om}^{2} ={}& 2 u^2 {\pd}_u\Gm_{0,\Om}^{2}+2 u v {\pd}_v\Gm_{0,\Om}^{2}+(q-2) u \Gm_{0,\Om}^{2}-2 u (q (u-v+1)-2 u+1) {\pd}_u\Gm_{1,\Om}^{2} \nn \\ 
		& +2 \left(q \left((v-1)^2-u (v+1)\right)+2 u v+u+v-1\right) {\pd}_v\Gm_{1,\Om}^{2}+4 u^2 {\pd}_u\Gm_{2,\Om}^{2} \nn \\ 
		& +2 u (2 v-3) {\pd}_v\Gm_{2,\Om}^{2}-4 u v (u-v+1) {\pd}_u {\pd}_v\Gm_{1,\Om}^{2} +2 u^2 (-u+v+1) {\pd}_u^2\Gm_{1,\Om}^{2} \nn \\ 
		& +2 v \left((v-1)^2-u (v+1)\right) {\pd}_v^2\Gm_{1,\Om}^{2} +\frac{2 (q-2) u (v-1) \Gm_{2,\Om}^{2}}{v}  \nn \\ 
		& +\left(-\frac{1}{6} q^2 (3 u-3 v+5)+q (3 u-v+2)-4 u+2\right) \Gm_{1,\Om}^{2}\,, \\
	\l\, \Gm_{2,\Om}^{2} ={}& -2 u v {\pd}_u\Gm_{0,\Om}^{2}-2 (v-1) v {\pd}_v\Gm_{0,\Om}^{2}-(q-2) v \Gm_{0,\Om}^{2} -4 u v {\pd}_u\Gm_{1,\Om}^{2} \nn \\ & +\left(2 q \left((v-1)^2-u (v+1)\right)+8 u-4 (v-2) (v-1)\right) {\pd}_v\Gm_{2,\Om}^{2} \nn \\ &-4 u v (u-v+1) {\pd}_u {\pd}_v\Gm_{2,\Om}^{2}-2 u (q (u-v+1)+2 v-3) {\pd}_u\Gm_{2,\Om}^{2} \nn \\ & +2 v \left((v-1)^2-u (v+1)\right) {\pd}_v^2\Gm_{2,\Om}^{2} +2 u^2 (-u+v+1) {\pd}_u^2\Gm_{2,\Om}^{2}\nn \\ & -2 (q-2) v \Gm_{1,\Om}^{2} +\left(-\frac{1}{6} q^2 (3 u+5)+\frac{2 (q-4) (u-1)}{v}+q (u+6) \right. \nn \\ & \left.+\frac{1}{2} (q-4) (q-2) v-12\right) \Gm_{2,\Om}^{2} +2 v (1-2 v) {\pd}_v\Gm_{1,\Om}^{2}  \,.
}

The three-point function between $\Jm$  and $\Phi$ is given by
\eq{
	\left \langle  \Phi(z_1) \, \Jm (z_2) \, \Om_I (z_3) \right \rangle ={}& \frac{1}{\bk{3}{1}^q \bk{3}{2} \bk{2}{3}} t_I \left( \Zbf_{3(21)} \right)\,.
}
Imposing \eqref{ChiralConstrain} and \eqref{CurrentConstrain}\footnote{In order to impose those constraints, we use \eqref{SCCD}, just as we did with the four-point functions.}, we find three possible operators that can be exchanged in the t-channel:
\seq{\label{PJ:ts}}{
	\Om_I\, :{}& &  {}&t_I \nn\\ 
	\Om_{(-\ell/2,q-\ell/2)}^{(\ell,\ell)} \, :{}& &  {}& \frac{\Xbf^{(\ell)}}{\left( \Xbf^2 \right)^{1+\ell}} \,, \label{PJ:t1}\\
	\Om_{\left((3+2\D-2q)/4,(-3+2\D-2q)/4\right)}^{(\ell,\ell+1)} \,:&& {}& \frac{\Tb^\ad \Xbf^{(\ell)}}{\left( \Xbf^2 \right)^{(\ell +q+3/2-\D)/2}} \,, \label{PJ:t2}\\
	\Om_{\left((3+2\D-2q)/4,(-3+2\D-2q)/4\right)}^{(\ell+1,\ell)} \,:&& {}& \frac{(\Xbf \Tb)^\a \Xbf^{(\ell)}}{\left( \Xbf^2 \right)^{(\ell +q+1/2-\D)/2}} \,. \label{PJ:t3}
}
Note that in \eqref{PJ:t1}, the dimension of the exchanged operator is $\D=q-\ell$, which for $\ell=0$ is a known short multiplet, $\Om=\bar\Phi$, and for $\ell>0$ the multiplet will be non-unitary. Regardless of this, we will solve the super Casimir for an operator with arbitrary dimension.

Just as before, we will look at the OPEs between $\phi \times J$, $\psi_\a \times J$ and $\phi \times j_\a$ and from there make an ansatz for the $\Am_1$, $\Bm_1$ and $\Bm_2$ functions. Taking the term proportional to $(\t_2\tb_4)$ in the LHS of \eqref{jjppb:4pf_tchannel}, we find
\eq{
	\left \langle J\, \psi_\a \, \phi^*\,\bar j_\ad \right \rangle ={}&4 a_0 \left[\left( \frac{\Bm_1}{u^{1/2}} +\frac{u^{1/2}}{v} \Bm_2\right) {\mathbb T}^4_{\a\ad} + \frac{u^{1/2}}{v} \Bm_2\, {\mathbb T}^2_{\a\ad} \right] \label{Bs_t_channel:ansatz} \,,
}
where \footnote{${\mathbb T}^i$ here should not be confused with the ones used in the s-channel.}
\eq{
	{\mathbb T}^1 \equiv \Km_4 {\mathbb I^{42}} \,, \qquad {\mathbb T}^2 \equiv \Km_4 {\mathbb I^{42}_{31}}  \,, \qquad \Km_4= \langle12 \rangle^{-\frac{q}{2}-\frac{3}{2}} \langle13 \rangle^{\frac{3}{2}-\frac{q}{2}} \langle14 \rangle^{q-2} \langle24 \rangle^{\frac{1}{2}-\frac{q}{2}} \langle34 \rangle^{-\frac{q}{2}-\frac{3}{2}} \,,
}
and $a_0=\frac{\ri}{e_2 \,e_4}$.

Now we proceed to analyze the $J \times \phi$, $j_\a \times \phi$ and $\psi \times J$ OPEs. For \eqref{PJ:t1}, since the three-point function has total $U(1)_R$ charge 0, the only terms appearing in the $J \times \phi$ are the lowest component of the long multiplet being exchange and its $\Qm\bar\Qm$ and $(\Qm\bar\Qm)^2$ descendants. For both $j_\a \times \phi$ and $\psi \times J$ OPEs, following the same argument, only the $\Qm$ and $\Qm^2\bar\Qm$ descendants of the lowest component of the multiplet can appear. Summarizing,
\seq{\label{PJ:t1OPE}}{
	J \times \phi \sim {}& A^{(\ell)}_{\D} + B^{(\ell+1)}_{\D+1} + C^{(\ell-1)}_{\D+1} + D^{(\ell)}_{\D+2} \,, \\
	j_\a \times \phi \sim {}& \xi^{(\ell+1,\ell)}_{\D+1/2} + \xi^{(\ell-1,\ell)}_{\D+1/2} + \chi^{(\ell,\ell+1)}_{\D+3/2} + \chi^{(\ell,\ell-1)}_{\D+3/2} \,, \\
						\sim {}& \psi \times J\,.
}

Both \eqref{PJ:t2} and \eqref{PJ:t3} have $U(1)_R$-charge one, therefore, in the $J \times \phi$ OPE we only find $\bar\Qm$ and $\bar \Qm^2$ descendants of the lowest component of the exchanged multiplet. Finally, for both $j_\a \times \phi$ and $\psi \times J$ OPEs, following the same argument, we find the that lowest component of the long multiplet being exchange and its $\Qm\bar\Qm$ and $(\Qm\bar\Qm)^2$ descendants appear in the OPE. Therefore, the OPEs from the three-point function \eqref{PJ:t2} are
\seq{\label{PJ:t2OPE}}{
	J \times \phi \sim {}& A^{(\ell+1)}_{\D+1/2} + B^{(\ell)}_{\D+3/2}\,, \\
	j_\a \times \phi \sim {}& \xi^{(\ell+1,\ell)}_{\D} + \chi^{(\ell+2,\ell+1)}_{\D+1} + \chi^{(\ell,\ell+1)}_{\D+1} + \chi^{(\ell,\ell-1)}_{\D+1} + \eta^{(\ell+1,\ell)}_{\D+2} \,,
}
and for \eqref{PJ:t3}, we find the following OPEs
\seq{\label{PJ:t3OPE}}{
	J \times \phi \sim {}& A^{(\ell)}_{\D+1/2} + B^{(\ell+1)}_{\D+3/2}\,, \\
	j_\a \times \phi \sim {}& \xi^{(\ell,\ell+1)}_{\D} + \chi^{(\ell+1,\ell+2)}_{\D+1} + \chi^{(\ell+1,\ell)}_{\D+1} + \chi^{(\ell-1,\ell)}_{\D+1} + \eta^{(\ell,\ell+1)}_{\D+2} \,,
}
with $\psi_\a \times J \sim j_\a \times \phi$. The reader should keep in mind that although $A$ and its $\Qm\bar\Qm$ descendants have the same $U(1)_R$-charge as the supermultiplet in \eqref{PJ:t1OPE}, this is no longer true in \eqref{PJ:t2OPE} and \eqref{PJ:t3OPE}.

Thus, in the case the exchange operator is $\Om_{(q-\ell/2,-\ell/2)}^{(\ell,\ell)}$, we have
\eq{
	\Gm_{0,\Om}^2(u,v) ={}& g_{\D,\ell}^{2-q,q-2} + c_1 g_{\D+1,\ell+1}^{2-q,q-2} + c_2 g_{\D+1,\ell-1}^{2-q,q-2} + c_3 g_{\D+2\ell}^{2-q,q-2} \,, \label{t_channel:At1}
}
and
\eq{
	4 a_0 \left[\left( \frac{\Gm_{1,\Om}^{2}}{u^{1/2}} +\frac{u^{1/2}}{v} \Gm_{2,\Om}^{2}\right) {\mathbb T}^4_{\a\ad} + \frac{u^{1/2}}{v} \Gm_{2,\Om}^{2}\, {\mathbb T}^2_{\a\ad} \right] ={}& c_4 W^{1,seed}_{\D+1/2,\ell,1} + c_5 W^{1,dual \,seed}_{\D+1/2,\ell-1,1} \nn \\ & + c_6 W^{1,dual \,seed}_{\D+3/2,\ell,1} + c_7 W^{1,seed}_{\D+3/2,\ell-1,1} \,, \label{t_channel:Bt1}
}
with
\seq{}{
	W^{1,seed}_{\D,\ell,1}={}& W^{seed}_{\langle J\psi_\a \Om_\D^{(\ell,\ell+1)} \rangle \langle \Om_\D^{(\ell+1,\ell)} \phi^* \bar j_\ad \rangle} \,, \\
	W^{1,dual\,seed}_{\D,\ell,1}={}& W^{seed}_{\langle J \psi_\a \Om_\D^{(\ell+1,\ell)} \rangle \langle \Om_\D^{(\ell,\ell+1)} \phi^* \bar j_\ad \rangle} \,.
}

Plugging this ansatz in \eqref{JJPPb:t_channel} gives us
\seq{\label{tc:01Sol}}{
	c_1 ={}& -\frac{(\D-q+\ell ) (\D+q+\ell -2)^2}{4 (\D+\ell ) (\D+\ell +1) (\D+q+\ell )} \,, \\
	c_2 ={}& -\frac{ (\D-q-\ell -2) (\D+q-\ell -4)^2}{4 (\D-\ell -2) (\D-\ell -1) (\D+q-\ell -2)} \,, \\
	c_3 ={}& \frac{ (\D-q-\ell -2) (\D+q-\ell -4)^2 (\D-q+\ell ) (\D+q+\ell -2)^2}{16 (\D-\ell -2) (\D-\ell -1) (\D+\ell ) (\D+\ell +1) (\D+q-\ell -2) (\D+q+\ell )} \,, \\
	c_4 ={}& -\frac{4 \ri a_0  (\D+q+\ell -2)}{\D+q+\ell } \,, \\
	c_5 ={}& \frac{4 \ri a_0  \ell  (-\D-q+\ell +4)}{(\ell +1) (-\D-q+\ell +2)} \,, \\
	c_6 ={}& -\frac{\ri a_0  \D (\D+q-\ell -4) (\D-q+\ell ) (\D+q+\ell -2)^2}{(\D-1) (\D+\ell ) (\D+\ell +1) (\D+q-\ell -2) (\D+q+\ell )} \,, \\
	c_7 ={}& \frac{\ri a_0  \D \ell  (\D-q-\ell -2) (\D+q-\ell -4)^2 (\D+q+\ell -2)}{(\D-1) (\ell +1) (\D-\ell -2) (\D-\ell -1) (\D+q-\ell -2) (\D+q+\ell )} \,, 
}

When exchanging an $\Om_{\left((-3+2\D-2q)/4,(3+2\D-2q)/4\right)}^{(\ell+1,\ell)}$ operator, we have
\eq{
	\Gm_{0,\Om}^{2} ={}& g_{\D+1/2,\ell+1}^{2-q,q-2} + c_1 g_{\D+3/2,\ell}^{2-q,q-2} \,, \label{t_channel:At2}
}
and
\eq{
	4 a_0 \left[\left( \frac{\Gm_{1,\Om}^{2}}{u^{1/2}} +\frac{u^{1/2}}{v} \Gm_{2,\Om}^{2}\right) {\mathbb T}^4_{\a\ad} + \frac{u^{1/2}}{v} \Gm_{2,\Om}^{2}\, {\mathbb T}^2_{\a\ad} \right] ={}& c_2 W^{1,seed}_{\D,\ell,1} + c_3 W^{1,seed}_{\D+1,\ell+1,1} \nn \\ & + c_4 W^{1,dual \,seed}_{\D+1,\ell,1} + c_5 W^{1,seed}_{\D+1,\ell-1,1} \nn \\ & + c_6 W^{1,seed}_{\D+2,\ell,1} \,. \label{t_channel:Bt2}
}
Plugging this ansatz in \eqref{JJPPb:t_channel} gives us
\seq{\label{tc:02Sol}}{
	c_1 ={}& -\frac{ (2 \D-3) (\ell +2) (2 \D-2 q-2 \ell -1) (2 \D+2 q-2 \ell -9)}{4 (2 \D-1) (\ell +1) (2 \D-2 \ell -5) (2 \D-2 \ell -3)} \,, \\
	c_2 ={}& -\ri a_0  (2 \D-2 q+2 \ell +3) \,, \\
	c_3 ={}& -\frac{\ri a_0  (2 \D-2 q+2 \ell +7) (2 \D+2 q+2 \ell -1)^2}{4 (2 \D+2 \ell +1) (2 \D+2 \ell +3)} \,, \\
	c_4 ={}& -\frac{\ri a_0  (2 \D-2 q-2 \ell -1) (2 \D+2 q-2 \ell -9) (4 \D \ell +6 \D-2 q-4 \ell -3)}{4 (2 \D-3) (2 \D-1) (\ell +1) (\ell +2)} \,, \\
	c_5 ={}& \frac{\ri a_0  \ell  (\ell +2) (2 \D-2 q-2 \ell -1) (2 \D+2 q-2 \ell -9) (2 \D-2 q+2 \ell +3)}{4 (\ell +1)^2 (2 \D-2 \ell -5) (2 \D-2 \ell -3)} \,, \\
	c_6 ={}& \frac{\ri a_0  (2 \D-3) (2 \D+1) (2 \D-2 q-2 \ell -1) (2 \D+2 q-2 \ell -9) (2 \D-2 q+2 \ell +7) }{16 (2 \D-1)^2 (2 \D-2 \ell -5) (2 \D-2 \ell -3) (2 \D+2 \ell +1) (2 \D+2 \ell +3)} \nn \\ & \qquad \times (2 \D+2 q+2 \ell -1)^2 \,.
}

Finally, for the exchange of an $\Om_{\left((-3+2\D-2q)/4,(3+2\D-2q)/4\right)}^{(\ell,\ell+1)}$ operator, we have
\eq{
	\Gm_{0,\Om}^{2} ={}& g_{\D+1/2,\ell}^{2-q,q-2} + c_1 g_{\D+3/2,\ell+1}^{2-q,q-2} \,, \label{t_channel:At3}
}
and
\eq{
	4 a_0 \left[\left( \frac{\Gm_{1,\Om}^{2}}{u^{1/2}} +\frac{u^{1/2}}{v} \Gm_{2,\Om}^{2}\right) {\mathbb T}^4_{\a\ad} + \frac{u^{1/2}}{v} \Gm_{2,\Om}^{2}\, {\mathbb T}^2_{\a\ad} \right] ={}& c_2 W^{1,dual\,seed}_{\D,\ell,1} + c_3 W^{1,dual\,seed}_{\D+1,\ell+1,1} \nn \\ & + c_4 W^{1,seed}_{\D+1,\ell,1} + c_5 W^{1,dual\,seed}_{\D+1,\ell-1,1} \nn \\ & + c_6 W^{1,dual\,seed}_{\D+2,\ell,1} \,. \label{t_channel:Bt3}
}
Plugging this ansatz in \eqref{JJPPb:t_channel} gives us
\seq{\label{tc:03Sol}}{
	c_1 ={}& -\frac{ (2 \D-3) (\ell +1) (2 \D-2 q+2 \ell +5) (2 \D+2 q+2 \ell -3)}{4 (2 \D-1) (\ell +2) (2 \D+2 \ell +1) (2 \D+2 \ell +3)} \,, \\
	c_2 ={}& -\frac{\ri a_0  (\ell +1) (-2 \D+2 q+2 \ell +3)}{\ell +2} \,, \\
	c_3 ={}& -\frac{\ri a_0  (\ell +1) (2 \D-2 q-2 \ell -3) (2 \D-2 q+2 \ell +5) (2 \D+2 q+2 \ell -3)}{4 (\ell +2) (2 \D+2 \ell +1) (2 \D+2 \ell +3)} \,, \\
	c_4 ={}& -\frac{\ri a_0  (2 \D-2 q+2 \ell +5) (2 \D+2 q+2 \ell -3) (4 \D \ell +6 \D+2 q-4 \ell -9)}{4 (2 \D-3) (2 \D-1) (\ell +2)^2} \,, \\
	c_5 ={}& \frac{\ri a_0  \ell  (2 \D-2 q-2 \ell +1) (2 \D+2 q-2 \ell -7)^2}{4 (\ell +1) (2 \D-2 \ell -5) (2 \D-2 \ell -3)} \,, \\
	c_6 ={}& -\frac{\ri a_0  (2 \D-3) (2 \D+1) (\ell +1) (2 \D-2 q-2 \ell +1) (2 \D+2 q-2 \ell -7)^2 }{16 (2 \D-1)^2 (\ell +2) (2 \D-2 \ell -5) (2 \D-2 \ell -3) (2 \D+2 \ell +1) (2 \D+2 \ell +3)} \nn \\ & \times (2 \D-2 q+2 \ell +5) (2 \D+2 q+2 \ell -3)\,. 
}

We point out that the superconformal blocks for the lowest components agree with the results found in \cite{Li:2017ddj}. Just as before, the normalization $a_0$ is irrelevant for the blocks of the lowest components, as expected. Again, we only needed to solve two eigenvalue equations and the third was identically satisfied. In the case of \eqref{PJ:t1OPE} we could have used either $\Gm_{1,\Om}^{2}$ or $\Gm_{2,\Om}^{2}$. In the case were the lowest component of the exchanged long multiplet was not present in the OPE, things were different: for \eqref{PJ:t2OPE}, the eigenequation for $\Gm_{0,\Om}^{2}$ we were able to fix all the $c_i$ coefficients but one, which was fixed using $\Gm_{2,\Om}^{2}$, while $\Gm_{1,\Om}^{2}$ did not give us any new information; the converse statement is true for \eqref{PJ:t3OPE}, where we used $\Gm_{1,\Om}^{2}$ to fix the remaining constant, while $\Gm_{2,\Om}^{2}$ was identically satisfied.

\subsection{\texorpdfstring{$\left\langle \Om \, \Om \, \Phi \, {\bar\Phi} \right\rangle$}{O O phi phidagger}}

So far, we have studied a special case where we only have short multiplets. As we saw in the previous section, we can replace the current multiplet and still have a simple four-point function. In \eqref{OOppb:4pf} and \eqref{OOppb:4pf_tchannel} we replaced the current multiplet for two arbitrary long multiplets such that the total $U(1)_R$-charge is zero. We will now work in the special case where $q_1=\bar q_2$ and $q_2=\bar q_1$. We will assume $q_1\neq \bar q_1$. The special case where the long multiplet has zero $U(1)_R$-charge is very similar to the case of two current multiplets and it was studied in detail in \cite{Li:2017ddj}. Since the lowest component of the long multiplet is a complex field, we cannot make a distinction between even and odd spin cases, as in \eqref{JJ:ansatz}. Therefore, the ansatz for the $\Am_1$ function will be very similar to the one given for the chiral case, see \eqref{PPbPPB:ansatz}. In the t-channel, we will study the same long multiplets being exchanged as in \eqref{PJ:ts} and we will obtain very similar results to \eqref{tc:01Sol}, \eqref{tc:02Sol} and \eqref{tc:03Sol}.

\subsubsection*{s-channel}

In the case where we replace the current multiplets by a long and its conjugated, \eqref{OOppb:4pf} now reads
\eq{
	\left\langle  \Om_{p,\bar p} (z_1)\, \Om_{\bar p,p}^{\dagger}(z_2)\, \Phi(z_3) \, {\bar\Phi}(z_4)\right\rangle ={}& \frac{1}{\bk{1}{2}^{{\bar p}} \bk{2}{1}^{p} \bk{4}{3}^{q}}  \left( \Am_1 + 4\ri \left [\left(222\right) - \left(422\right) \right]\Bm_1 \right. \nn \\
	& \qquad  \left. + 4\ri \left [\left(242\right) - \left(442\right) \right]\Bm_2 + \left[ (222)^2 \right. \right. \nn \\ & \qquad \left.\left. +2(222)(422)  +(422)^2\right] \Cm_1 \right)\,.
}

First, we expand the correlator
\seq{}{
	\Am_1 ={}& \sum_\Om \Gm_{0,\Om}^{3}(u,v) \,, \\
	\Bm_1 ={}& \sum_\Om \Gm_{1,\Om}^{3}(u,v) \,, \\
	\Bm_2 ={}& \sum_\Om \Gm_{2,\Om}^{3}(u,v) \,, \\
	\Cm_1 ={}& \sum_\Om \Gm_{3,\Om}^{3}(u,v) \,.
}

Acting with super Casimir \eqref{sCas} on the chirals we find four eigen equations,
\seq{\label{OOPPb:CasEq}}{
	\l \,\Gm_{0,\Om}^{3} = {}& 2 u^2 (-u+v+1) {\pd}_u^2  \Gm_{0,\Om}^{1} -2 \left(2 u v+u-2 v^2+3 v-1\right) {\pd}_v  \Gm_{0,\Om}^{1} \nn \\ &-4 u (u-v+1)  {\pd}_u  \Gm_{0,\Om}^{1}  -4 u v (u-v+1) {\pd}_u {\pd}_v  \Gm_{0,\Om}^{1}+2 v \left((v-1)^2 \right. \nn \\ & \left.-u (v+1)\right) {\pd}_v^2  \Gm_{0,\Om}^{1} +\frac{2 (u+v-1) \Gm_{1,\Om}^{1} }{v} +4 \Gm_{2,\Om}^{1} \,,     \\
	\l \,\Gm_{1,\Om}^{3} = {}& 2 u^2 (-u+v+1) {\pd}_u^2  \Gm_{1,\Om}^{1} -4 u^2 {\pd}_u  \Gm_{1,\Om}^{1} -4 u v {\pd}_u  \Gm_{2,\Om}^{1} -4 u v (u-v+1) {\pd}_u {\pd}_v  \Gm_{1,\Om}^{1} \nn \\ & +2 (-2 u v+u+v-1) {\pd}_v  \Gm_{1,\Om}^{1} +2 v \left((v-1)^2-u (v+1)\right) {\pd}_v^2  \Gm_{1,\Om}^{1} -4 v^2 {\pd}_v  \Gm_{2,\Om}^{1} \nn \\ & -\frac{2 (u+v-1) \Gm_{1,\Om}^{1} }{v} \,,     \\
	\l \,\Gm_{2,\Om}^{3} = {}& 2 u^2 (-u+v+1) {\pd}_u^2  \Gm_{1,\Om}^{1} -4 u^2 {\pd}_u  \Gm_{1,\Om}^{1} -4 u v {\pd}_u  \Gm_{2,\Om}^{1} -4 u v (u-v+1) {\pd}_u {\pd}_v  \Gm_{1,\Om}^{1} \nn \\ & +2 (-2 u v+u+v-1) {\pd}_v  \Gm_{1,\Om}^{1} +2 v \left((v-1)^2-u (v+1)\right) {\pd}_v^2  \Gm_{1,\Om}^{1} -4 v^2 {\pd}_v  \Gm_{2,\Om}^{1} \nn \\ & -\frac{2 (u+v-1) \Gm_{1,\Om}^{1} }{v} -\frac{4}{4v}\Gm_{3,\Om}^{3}\, , \\
	\l\, \Gm_{3,\Om}^{3} = {}& 2 u^2 (-u+v+1)\pd_u^2 \Gm_{3,\Om}^{3} -4 u v (u-v+1) \pd_u\pd_v\Gm_{3,\Om}^{3} -\frac{2 (u+v-1)}{v}\Gm_{3,\Om}^{3}  \nn \\ & +2 (u+v-1) \pd_v\Gm_{3,\Om}^{3} +2 v \left((v-1)^2-u (v+1)\right) \pd_v^2\Gm_{3,\Om}^{3} \,. \label{OOPPb:CasEq_4}
}
Note that the eigen equations for $\Gm_{0,\Om}^{3}$ and $\Gm_{1,\Om}^{3}$ are the same as in \eqref{JJPPb:CasEq}. The eigen equation for $\Gm_{2,\Om}^{3}$ has an extra $-\frac{4}{4v}\Gm_{3,\Om}^{3}$ term compared to the one in \eqref{JJPPb:CasEq}.

Again, we look at the three-point functions in order to make an ansatz for $\Am_1$ and the $\Bm_i$ functions, and also for $\Cm_1$. We know that the only contribution to the super blocks comes from a $\Om_{(\D/2,\D/2)}^{(\ell)}$ multiplet \eqref{PPb:3pf}. The three-point function between the longs and such operator was already written in the introduction,
\eq{
	\left \langle \Om_{(p,\bar p)}(z_1) \, \Om_{(\bar p,p)}^{\dagger}(z_2) \, \Om_{(\D/2,\D/2)}^{(\ell)}(z_3) \right \rangle ={}& \frac{1}{\bk{1}{3}^{\bar p} \bk{3}{1}^p \bk{2}{3}^p \bk{3}{2}^{\bar p}} \frac{1}{\left( \Xbf \cdot \Xbfb\right)^{p+\bar p - (\D-\ell)/2}} \nn \\
	& \times \left[ \l_+^{(1)} t_{1+}^{(\ell)} + \l_-^{(2)} t_{2-}^{(\ell)} + \l_-^{(3)} t_{3-}^{(\ell)} + \l_4^{(4)} t_{4+}^{(\ell)}  \right] \,, \label{3pf:OOO}
}
with
\seq{\label{ts:longs}}{
	t_{1+}^{(\ell)}={}& \Xbf_+^{(\ell)}\,,	 &	 t_{4+}^{(\ell)}={}& \Xbf_+^{(\ell)} \frac{\T^2\Tb^2}{\Xbf\cdot\Xbfb}\,,\\
	t_{2-}^{(\ell)}={}& \Xbf_+^{(\ell-1}\Xbf_-^{1)}\,,	&	 t_{3-}^{(\ell)}={}& \Xbf_+^{(\ell)} \frac{\Xbf_+\cdot\Xbf_-}{\Xbf\cdot\Xbfb}\,. 
}
The subscript $t_{i\pm}^{\ell}$ means that under $z_1\leftrightarrow z_2$, $t_{i\pm}^{\ell} \to \pm (-)^\ell t_{i\pm}^{\ell}$. The structures appearing in \eqref{3pf:OOO} are the same as the ones appearing in \eqref{3pf:JJO}, but since there are no constraints that can be imposed to the long multiplets, there is no equation relating them.

In order to get the $m_{p+\bar p}\times m_{p+\bar p}^*$ OPE, being $m_{p+\bar p}$ the lowest component of the long multiplets, see \eqref{multiplets_expansion}, we perform a similar analysis to the ones we have carried out so far. Since the $t_{1+}^{(\ell)}$ structure does not vanish when taking the fermionic coordinates to zero, and the three-point function \eqref{3pf:OOO} has zero $U(1)_R$-charge, its contribution to the OPE will have the lowest operator of the exchanged long multiplet, its $\Qm\bar\Qm$ and its $(\Qm\bar\Qm)^2$ descendants. Both $t_{2-}^{(\ell)}$ and $t_{3-}^{(\ell)}$ are proportional to $\T_{3(12)}\,\Tb_{3(12)}$, thus, they will only contribute with the $\Qm\bar\Qm$ and $(\Qm\bar\Qm)^2$ descendants of the lowest operator of the exchanged multiplet to the OPE. Finally, $t_{4+}^{(\ell)}$ is proportional to  $(\T_{3(12)}\,\Tb_{3(12)})^2$, therefore, its only contribution to the OPE is the $(\Qm\bar\Qm)^2$ descendant of the lowest component of the exchanged multiplet. Summarizing, the OPE between the lowest component of the long multiplets is
\eq{
	m_{p+\bar p}\times m_{p+\bar p}^* \sim{} A_{\D}^{(\ell)} +B_{\D+1}^{(\ell+1)} + C_{\D+1}^{(\ell-1)} + D_{\D+2}^{(\ell)}\,,
}
which is of the same form as the $\phi \times \phi ^*$ OPE, \eqref{OPE:ppb}. Thus, the ansatz for $\Am_1$ is given by \eqref{PPbPPB:ansatz}:
\eq{
	\Gm_{0,\Om}^{3}(u,v)={}& c_1g_{\D,\ell} + c_2 g_{\D+1,\ell+1} + c_3 g_{\D+1,\ell-1} + c_4 g_{\D+2,\ell} \,,  \label{OOPPB:ansatz}
}
but now we note that, unlike the $\langle \Phi\,\Phi^\dagger\Phi\,\Phi^\dagger\rangle$ the $c_i$ coefficients depends on several three-point functions coefficients.  While $c_1$ only depends on $\l_+^{(1)}$, while both $c_2$ and $c_3$ are functions of $\l_+^{(1)}$, $\l_-^{(2)}$ and $\l_-^{(3)}$, and finally $c_4$ will depend on all the three-point function coefficients \cite{Khandker:2014mpa,Li:2016chh}, following our previous discussion.

For the $\Bm_i$ functions, we again look at the term proportional to $\t_2\bar\t_4$. This turns out to be again \eqref{4pf1}, replacing the dimension fo the current multiplet by $p+\bar p$. The $m_{p+\bar p} \times n_{(p+\bar p)\a}^\dagger$ OPE, being $n_{(p+\bar p)\ad}$ the first $\bar\Qm$ descendant of the long multiplet, is of the same form as the $J\times j_\a$ OPE, \eqref{OPE:Jj}: studying the terms proportional to $\t_2$ in the three-point function \eqref{3pf:OOO}, we find that the $t_{1+}^{(\ell)}$, $t_{2-}^{(\ell)}$ and $t_{3-}^{(\ell)}$ structures contribute with the $\bar \Qm$ and $\Qm\bar\Qm^2$ descendants of the lowest component of the exchanged multiplet, while $t_{4+}^{(\ell)}$ only contributes with the $\Qm\bar\Qm^2$ descendants of the lowest component of the exchanged multiplet. Therefore, the required OPE is
\eq{
	m_{p+\bar p} \times n_{(p+\bar p)\a}^\dagger \sim{} & \xi_{\D+1/2}^{(\ell+1,\ell)} + \xi_{\D+1/2}^{(\ell-1,\ell)} + \chi_{\D+3/2}^{(\ell,\ell+1)} + \chi_{\D+3/2}^{(\ell,\ell-1)} \,. \label{OPE:mn}
}

Thus, our ansatz for $\Gm_{1,\Om}^{3}$ and $\Gm_{2,\Om}^{3}$ is now given by
\eq{
	4 a_0\left[\left( \frac{1}{u^{1/2}}\Gm_{2,\Om}^{3} + \frac{u^{1/2}}{v} \Gm_{1,\Om}^{3} \right){\mathbb T}^4_{\a\ad} + \frac{u^{1/2}}{v} \Gm_{1,\Om}^{3} {\mathbb T}^2_{\a\ad} \right] ={}& c_5 W^{2,seed}_{\D+1/2,\ell,1} + c_6 W^{2,seed}_{\D+3/2,\ell-1,1} \nn \\ &  + c_7 W^{2,dual\,seed}_{\D+1/2,\ell-1,1} + c_8 W^{2,dual\,seed}_{\D+3/2,\ell,1}\,. \label{Bs:ansats_OOPPB}
}
with $a_0=\frac{\ri}{e_8 \,e_6}$ and
\seq{}{
	W^{2,seed}_{\D,\ell,1}={}& W^{seed}_{\langle m_{p+\bar p} r^\dagger_{p+\bar p+1/2,\a} \Om_\D^{(\ell,\ell+1)} \rangle \langle \Om_\D^{(\ell+1,\ell)} \phi \bar\psi_\ad \rangle} \,, \\
	W^{2,dual\,seed}_{\D,\ell,1}={}& W^{seed}_{\langle m_{p+\bar p} r^\dagger_{p+\bar p+1/2,\a} \Om_\D^{(\ell+1,\ell)} \rangle \langle \Om_\D^{(\ell,\ell+1)} \phi \bar\psi_\ad \rangle} \,.
}

Finally, we need to give an ansatz for $\Gm_{3,\Om}^3$, which can be done by looking at the term proportional to $\t_2^2 \tb_4^2$ in the four-point function:
\eq{
	\left \langle m_{p+\bar p}\, o_{p+\bar p+1}\, \phi\, F^* \right \rangle ={} & -\frac{1}{2} a_1\Km_6  \frac{u^{1/2}}{v}\Gm_{3,\Om}^3\,, \label{Cm:ansatz_s}
}
with $a_1=\frac{1}{e_{10}e_7}$ and
\eq{
	\Km_6 ={}& \frac{1}{\bkb{1}{2}^{p+\bar p}\bkb{3}{4}^{q}} \left(\frac{\bkb{1}{3}}{\bkb{1}{2}\bkb{3}{4}\bkb{2}{4}}\right)^{1/2} \,.
}

In order to give an ansatz for \eqref{Cm:ansatz_s}, we need to look at the $m_{p+\bar p} \times o_{p+\bar p +1}$ and $\phi \times F^*$ OPEs. These OPEs can be found by looking at the terms proportional to $\t_2^2$ in \eqref{3pf:OOO} and to $\t_1^2$ in \eqref{PPb:3pf}. The only term appearing in those expansions is proportional to $\tb_3^2$, which corresponds to a ${\bar \Qm}^2$ descendant of the $\Om^{(\ell)}$ long multiplet. Thus, our OPEs are
\eq{
	m_{p+\bar p} \times o_{p+\bar p +1} \sim E^{\ell}_{\D+1}
}
and
\eq{
	\phi \times F^* \sim E^{\ell}_{\D+1}\,.
}
Our ansatz for $\Gm_{3,\Om}^3$ is then given by
\eq{
	-\frac{1}{2} a_1 \frac{u^{1/2}}{v}\Gm_{3,\Om}^3 ={}& c_9 g_{\D+1,\ell}\,,
}
Note that this ansatz satisfies \eqref{OOPPb:CasEq_4} identically.

Solving the eigenvalue equations with these ansatz, we obtain
\seq{\label{sc:Long}}{
	c_5= {}& \ri a_0 c_1 (\Delta +\ell )+4 \ri a_0 c_2 (\Delta +\ell +1)\,, \\ 
	c_6= {}& \frac{\ri a_0 c_3 \Delta  \ell  (\Delta +\ell )}{(\Delta -1) (\ell +1)}+\frac{4 \ri a_0 c_4 \Delta  \ell  (\Delta +\ell +1)}{(\Delta -1) (\ell +1)}\,, \\ 
	c_7= {}& \frac{\ri a_0 c_1 \ell  (-\Delta +\ell +2)}{\ell +1}+\frac{4 \ri a_0 c_3 \ell  (-\Delta +\ell +1)}{\ell +1}\,, \\ 
	c_8= {}& -\frac{\ri a_0 c_2 \Delta  (\Delta -\ell -2)}{\Delta -1}-\frac{4 \ri a_0 c_4 \Delta  (\Delta -\ell -1)}{\Delta -1}\,, \\ 
	c_9= {}& \frac{a_1 c_1 (\Delta -2) (\Delta -\ell -2) (\Delta +\ell )}{4 (\Delta -1)}+\frac{4 a_1 c_4 \Delta  (\Delta -\ell -1) (\Delta +\ell +1)}{\Delta -1} \nn \\ & -\frac{a_1 c_2 (\ell +2) (-\Delta +\ell +2) (\Delta +\ell +1)}{\ell +1}-\frac{a_1 c_3 \ell  (-\Delta +\ell +1) (\Delta +\ell )}{\ell +1} \,,
}
while leaving $c_1$, $c_2$, $c_3$ and $c_4$ unfixed. In order to fix those coefficients, one can, for example, use the supershadow formalism \cite{Li:2016chh}.

\subsubsection*{t-channel}

For the t-channel, the four-point function \eqref{OOppb:4pf_tchannel} reads
\eq{
	\left\langle \Om_{(p,\bar p)}(z_1) \, \Phi(z_2) \, {\bar\Phi}(z_3) \, \Om^{\dagger}_{(\bar p,p)}(z_4) \right\rangle ={}& \frac{u^{-\frac{q+p+\bar p}{2}} v^q}{\bk{1}{4}^{\bar p} \bk{4}{1}^{p} \bk{3}{2}^q } \left(\Am_1  + 4\ri \left [\left(111\right) - \left(311\right) \right]\Bm_1 \right. \nn \\
	& \left.+ 4\ri \left [\left(131\right) - \left(331\right) \right]\Bm_2 +  (111)^2\Cm_1 + (311)^2 \Cm_1 \right. \nn \\
	& \left.+2(111)(311)\Cm_1 \right)
}
Expanding the correlator\footnote{Unfortunately, we have run out of letters for the exchanged long multiplet.}
\seq{\label{spw:expansion}}{
	\Am_1 ={}& \sum_\Om \Gm_{0,\tilde\Om}^{4}(u,v) \,, \\
	\Bm_1 ={}& \sum_\Om \Gm_{1,\tilde\Om}^{4}(u,v) \,, \\
	\Bm_2 ={}& \sum_\Om \Gm_{2,\tilde\Om}^{4}(u,v) \,, \\
	\Cm_1 ={}& \sum_\Om \Gm_{3,\tilde\Om}^{4}(u,v) \,.
}

The super Casimir equations are now more complicated than the ones involving the current multiplet \eqref{JJPPb:t_channel}. Applying \eqref{sCas} we obtain four eigenvalue equations
\seq{\label{OOPPb:t_channel}}{
	\l\, \Gm_{0,\tilde\Om}^{4} ={}& 2 u^2 (u-v-1) \pd_u\Gm_{2,\tilde\Om}^{4}+2 u v (u-v+1) \pd_u\Gm_{1,\tilde\Om}^{4}+2 u v (u-v+1) \pd_v\Gm_{2,\tilde\Om}^{4} \nn \\ 
		& +2 v \left(u (v+1)-(v-1)^2\right) \pd_v\Gm_{1,\tilde\Om}^{4}+u \Gm_{2,\tilde\Om}^{4} ((p+\bar p-q) (-u+v+1)+4) \nn \\ 
		& -v (u-v+1) (p+{\bar p}-q) \Gm_{1,\tilde\Om}^{4}\,, \\
	\l\, \Gm_{1,\tilde\Om}^{4} ={}& \pd_v\Gm_{0,\tilde\Om}^{4} \left(\left(2 u (v+1)-2 (v-1)^2\right) (p+{\bar p}-q+1)-2 u (3 v+2)+6 v^2-10 v+4\right) \nn \\ 
		& +2 u (u-v+1) (p+{\bar p}-q-2) \pd_u\Gm_{0,\tilde\Om}^{4}+2 u^2 (-u+v+1) \pd_u^2\Gm_{0,\tilde\Om}^{4} \nn \\ 
		& -4 u v (u-v+1) \pd_u {\pd}_v\Gm_{0,\tilde\Om}^{4}+2 v \left((v-1)^2-u (v+1)\right) \pd_v^2\Gm_{0,\tilde\Om}^{4} \nn \\ 
		& -\frac{2 (u+v-1) \Gm_{2,\tilde\Om}^{4}}{v} -4 \Gm_{1,\tilde\Om}^{4} +\Gm_{0,\tilde\Om}^{4} \left(+\frac{2}{3} (p+{\bar p}-q+1) (2 p+3 u-3 v+4)\right. \nn \\ 
		& \left.(-3 u+3 v-5) (p+{\bar p}-q+1)^2+\frac{1}{6} (-8 p (p+1)-9 u+9 v-11) \right) +\frac{1}{4} \Gm_{3,\tilde\Om}^{4}  \,, \\
	\l\, \Gm_{2,\tilde\Om}^{4} ={}& 2 p u^2 \pd_u\Gm_{0,\tilde\Om}^{4}+2 p u v \pd_v\Gm_{0,\tilde\Om}^{4}-p u (p+{\bar p}-q) \Gm_{0,\tilde\Om}^{4}+2 u^2 (-u+v+1) \pd_u^2\Gm_{1,\tilde\Om}^{4} \nn \\ 
		& +\pd_v\Gm_{1,\tilde\Om}^{4} \left( \left(2 u (v+1)-2 (v-1)^2\right) (p+{\bar p}-q+1)-2 u (v+2) \right. \nn \\ 
		& \left. +6 v^2-10 v+4 \right) +2 u \pd_u\Gm_{1,\tilde\Om}^{4} ((u-v+1) (p+{\bar p}-q+1)-  (u-3 v+4) ) \nn \\ 
		& +4 u^2 \pd_u\Gm_{2,\tilde\Om}^{4} +2 u (2 v-3) \pd_v\Gm_{2,\tilde\Om}^{4}-4 u v (u-v+1) \pd_u {\pd}_v\Gm_{1,\tilde\Om}^{4} \nn \\ 
		& +2 v \left((v-1)^2-u (v+1)\right) \pd_v^2\Gm_{1,\tilde\Om}^{4} +\Gm_{1,\tilde\Om}^{4} \left(\frac{1}{6} \left(-8 p^2+4 p+3 u+9 v+7\right) \right. \nn \\ 
		& \left. +\frac{1}{6} (-3 u+3 v-5) (p+{\bar p}-q+1)^2 +\frac{1}{3} (4 p-6 v+5) (p+{\bar p}-q+1)\right) \nn \\
		& -\frac{2 u (v-1) (p+{\bar p}-q) \Gm_{2,\tilde\Om}^{4}}{v} \,,
}
\eq{
	\l\, \Gm_{3,\tilde\Om}^{4} ={}& -\frac{16}{v} (p-1) u^2 (u-v-1)  {\pd}_u{\Bm}_2 -16 (p-1) u (u-v+1)  {\pd}_u {\Bm}_1 \nn \\
		& -16 (p-1) u (u-v+1)  {\pd}_v {\Bm}_2 -16 (p-1) \left(u (v+1)-(v-1)^2\right)  {\pd}_v{\Bm}_1 \nn \\
		& -\frac{8u }{v}(p-1)((p+{\bar p}) (-u+v+1)+q (u-v-1)+4){\Bm}_2 \nn \\
		&+8 (p-1) (u-v+1) (p+{\bar p}-q) {\Bm}_1 +2 u^2 (-u+v+1)  {\pd}_u^2{\Cm}_1 \nn \\
		&-2 u  ((p+{\bar p}) (-u+v-1)+q (u-v+1)+2) {\pd}_u{\Cm}_1 \nn \\
		& -2 \left(u (v+1)-(v-1)^2\right) (-p-{\bar p}+q)  {\pd}_v{\Cm}_1-4 u v (u-v+1)  {\pd}_u {\pd}_v{\Cm}_1 \nn \\
		& +2 v \left((v-1)^2-u (v+1)\right)  {\pd}_v^2{\Cm}_1 +\frac{1}{6} {\Cm}_1(u,v) \left(-4 q (p-{\bar p}) +q^2 (-3 u+3 v-5) \right. \nn \\
		& -3 (p+{\bar p}) (p+{\bar p}+2) (u-v+1)-2 (p-{\bar p}-6) (p-{\bar p})\nn \\
		& \left. + 6 q ((p+{\bar p}) (u-v+1)+u-v+3)\right) \label{OOPPb:t_channel_1}
}

The next step can be guessed by the reader: we will look at the three-point functions and see which operators are being exchanged. Most of the work was already done in \eqref{PJ:ts}, where we note that now \eqref{PJ:t1} does not have a fixed power in the denominator, and since there is no conservation equation for the long multiplet, there is another structure. The three-point function is 
\eq{
	\left \langle  \Phi(z_1) \, \tilde\Om_{(p,\bar p)} (z_2) \, \Om_I (z_3) \right \rangle ={}& \frac{1}{\bk{3}{1}^q \bk{3}{2}^p \bk{2}{3}^{\bar p}} t_I \left( \Zbf_{3(21)} \right)\,.
}
The long multiplet version of \eqref{PJ:ts} reads
\seq{\label{PO:ts}}{
	\Om_I\, :{}& &  {}&t_I \nn\\
	\tilde\Om_{( (\D-p+{\bar p}-q)/2, (\D+p-{\bar p}+q)/2)}^{(\ell,\ell)} \, :{}& &  & \frac{\Xbf^{(\ell)}}{\left( \Xbf^2 \right)^{(-\D+p+{\bar p}+q+\ell)/2}} \,, \label{PO:t1}\\
	\tilde\Om_{( \left(\D-p+{\bar p}-q+3/2\right)/2, \left(\D+p-{\bar p}+q-3/2\right)/2)}^{(\ell,\ell+1)} \,:&& & \frac{\Tb^\ad \Xbf^{(\ell)}}{\left( \Xbf^2 \right)^{(-\D+p+{\bar p}+q+\ell +1/2)/2}} \,, \label{PO:t2}\\
	\tilde\Om_{( \left(\D-p+{\bar p}-q+3/2\right)/2, \left(\D+p-{\bar p}+q-3/2\right)/2)}^{(\ell+1,\ell)} \,:&& & \frac{(\Xbf \Tb)^\a \Xbf^{(\ell)}}{\left( \Xbf^2 \right)^{(-\D+p+{\bar p}+q+\ell +3/2)/2}} \,. \label{PO:t3}
}

We will make an ansatz for $\Gm_{0,\tilde\Om}^{4}$ by inspecting the OPEs between the lowest component of the multiplets of $\Om$ and $\Phi$. The ansatz for the $\Gm_{1,\tilde\Om}^{4}$ and $\Gm_{2,\tilde\Om}^{4}$ functions will come from the OPEs of the $\Qm$-descendants of the lowest component of the multiplets and \eqref{Bs_t_channel:ansatz}, where we only need to change the dimension of the current multiplet for the dimension of the long multiplet. Since \eqref{PJ:t1}, \eqref{PJ:t2} and \eqref{PJ:t3} have the same structures as \eqref{PO:t1}, \eqref{PO:t2} and \eqref{PO:t3}, respectively, the OPEs for those cases are the same as the ones given in \eqref{PJ:t1OPE}, \eqref{PJ:t2OPE} and \eqref{PJ:t2OPE}, and we only need to replace the $J$ and its $\Qm$ and $\bar\Qm$ descendants by $m_{p+\bar p}$ and its $\Qm$ and $\bar\Qm$ descendants. Finally the ansatz for $\Gm_{3,\tilde\Om}^{4}$ will come from the $\Qm^2$-descendants of the lowest components of the multiplets, as we saw previously in the s-channel case.

When exchanging an $\tilde\Om_{( (\D+p-{\bar p}+q)/2,(\D-p+{\bar p}-q)/2)}^{(\ell,\ell)}$ long multiplet, the ansatz are similar to \eqref{t_channel:At1}  and \eqref{t_channel:Bt1}, and we have
\eq{
	\Gm_{0,\tilde\Om}^{4} ={}& g_{\D,\ell}^{p+{\bar p}-q,q-p-{\bar p}} + c_1 g_{\D+1,\ell+1}^{p+{\bar p}-q,q-p-{\bar p}} + c_2 g_{\D+1,\ell-1}^{p+{\bar p}-q,q-p-{\bar p}} + c_3 g_{\D+2\ell}^{p+{\bar p}-q,q-p-{\bar p}} \,, \label{t_channel:At1Long}
} 
and
\eq{
	4 a_0 \left[\left( \frac{\Gm_{1,\tilde\Om}^{4}}{u^{1/2}} +\frac{u^{1/2}}{v} \Gm_{2,\tilde\Om}^{4}\right) {\mathbb T}^4_{\a\ad} + \frac{u^{1/2}}{v} \Gm_{2,\tilde\Om}^{4}\, {\mathbb T}^2_{\a\ad} \right] ={}& c_4 W^{3,seed}_{\D+1/2,\ell,1} + c_5 W^{3,dual \,seed}_{\D+1/2,\ell-1,1} \nn \\ & + c_6 W^{3,dual \,seed}_{\D+3/2,\ell,1} + c_7 W^{3,seed}_{\D+3/2,\ell-1,1} \,. \label{t_channel:Bt1Long}
}
with
\seq{}{
	W^{3,seed}_{\D,\ell,1}={}& W^{seed}_{\langle m_{p+\bar p}\psi_\a \Om_\D^{(\ell,\ell+1)} \rangle \langle \Om_\D^{(\ell+1,\ell)} \phi^* n^\dagger_{p+\bar p+1/2,\ad} \rangle} \,, \\
	W^{3,dual\,seed}_{\D,\ell,1}={}& W^{seed}_{\langle m_{p+\bar p} \Om_\D^{(\ell+1,\ell)} \rangle \langle \Om_\D^{(\ell,\ell+1)} \phi^* n^\dagger_{p+\bar p+1/2,\ad} \rangle} \,,
}
and $a_0=\frac{\ri}{e_4\,e_9}$.

Finally, in order to find an ansatz for $\Gm_{3,\tilde\Om}^{4}$, we look at the term proportional to $\t_2^2\,\tb_4^2$ in the four-point function:
\eq{
	\left\langle m_{p+\bar p}\, F\,\phi^*\,o_{p+\bar p}^* \right \rangle ={}& -\frac{1}{2}a_1 \Km_7 \frac{1}{u^{1/2}}\Gm_{3,\tilde\Om}^{4}\,,
}
with $a_0=\frac{1}{e_5 \, e_{11}}$ and
\eq{
	\Km_7={}& \left(\frac{\bkb{1}{4}}{\bkb{1}{3}}\right)^q \left(\frac{\bkb{2}{4}}{\bkb{1}{4}}\right)^{p+\bar p} \left(\frac{\bkb{1}{3}}{\bkb{1}{2}\bkb{2}{4}\bkb{3}{4}}\right)^{(p+\bar p+q+1)/2} \,.
}

To find the $m_{p+\bar p}\times F$ and $o_{p+\bar p+1} \times \phi$ OPEs is similar to the s-channel case. By taking \eqref{PO:t1} $\t_1=\tb_1=\tb_2=0$ and looking at the terms proportional to $\t_2^2$ we only find a $\tb_3^2$ term, which corresponds to a ${\bar \Qm}^2$ descendant of the lowest component of the exchanged multiplet. For the second OPE we look at the term proportional to  $\t_1^2$ while taking $\tb_1=\t_2=\tb_2=0$, which again we only find a $\tb_3^2$ term which corresponds to the same descendant. Therefore
\eq{
	m_{p+\bar p}\times F \sim E^{\ell}_{\D+1}\,, \\
	o_{p+\bar p+1}\times \phi \sim E^{\ell}_{\D+1}\,.
}
Our ansatz for $\Gm_{3,\tilde\Om}^{4}$ now reads
\eq{
	-\frac{1}{2}a_1 \frac{1}{u^{1/2}}\Gm_{3,\tilde\Om}^{4}=c_8 g_{\D+1,\ell}^{p+\bar p -q-1,q-p-\bar p -1}\,.
}

Using the super Casimir equations we find
\seq{\label{tc2:01Sol}}{
	c_1 ={}& -\frac{1}{4} m_{0,2}\, m_{0,3} \, \\
	c_2 ={}& \frac{1}{4}  m_{0,1}\, m_{0,4} \, \\
	c_3 ={}& c_1 \,c_2 \, \\
	c_4 ={}& -4 {\ri}\, a_0\, {p}\, m_{0,2} \, \\
	c_5 ={}& \frac{4 {\ri}\, a_0\, {p}\, \ell \, m_{0,1}}{\ell +1} \, \\
	c_6 ={}& \frac{4 {\ri}\, a_0 \,\Delta \, {p} \,m_{0,1} }{\Delta -1}c_1 \, \\
	c_7 ={}& -\frac{4 {\ri}\, a_0\, \Delta\,  {p}\, \ell  \,m_{0,2}}{(\Delta -1) (\ell +1)}c_2 \,, \\
	c_8 ={}& 4 a_1 ({p}-1) \,{p}\, m_{0,1}\, m_{0,2} 
}
with
\seq{}{
	m_{0,1}={}& -\frac{-\Delta +p+{\bar p}-q+\ell +2}{\Delta +p-{\bar p}+q-\ell -2}\, \\ 
	m_{0,2}={}& -\frac{-\Delta +p+{\bar p}-q-\ell }{\Delta +p-{\bar p}+q+\ell }\, \\ 
	m_{0,3}={}& \frac{(\Delta -p+{\bar p}-q+\ell ) (\Delta -p-{\bar p}+q+\ell )}{(\Delta +\ell ) (\Delta +\ell +1)}\, \\ 
	m_{0,4}={}& -\frac{(-\Delta +p+{\bar p}-q+\ell +2) (-\Delta +p-{\bar p}+q+\ell +2)}{(-\Delta +\ell +1) (-\Delta +\ell +2)}
}

When the exchanged operator is $\tilde\Om_{( \left(\D+p-{\bar p}+q-3/2\right)/2,\left(\D-p+{\bar p}-q+3/2\right)/2)}^{(\ell+1,\ell)}$ the ansatz for $\Gm_{0,\tilde\Om}^{4}$ corresponds to, see \eqref{t_channel:At2}, 
\eq{
	\Gm_{0,\tilde\Om}^{4} ={}& g_{\D+1/2,\ell+1}^{2-q,q-2} + c_1 g_{\D+3/2,\ell}^{2-q,q-2} \,, \label{t_channel:At2Long}
}
while the anstaz for $\Gm_{1,\tilde\Om}^{4}$ and $\Gm_{2,\tilde\Om}^{4}$ is given by, see \eqref{t_channel:Bt2},
\eq{
	4 a_0 \left[\left( \frac{\Gm_{1,\tilde\Om}^{4}}{u^{1/2}} +\frac{u^{1/2}}{v} \Gm_{2,\tilde\Om}^{4}\right) {\mathbb T}^4_{\a\ad} + \frac{u^{1/2}}{v} \Gm_{2,\tilde\Om}^{4}\, {\mathbb T}^2_{\a\ad} \right] ={}& c_2 W^{3,seed}_{\D,\ell,1} + c_3 W^{3,seed}_{\D+1,\ell+1,1} \nn \\ & + c_4 W^{3,dual \,seed}_{\D+1,\ell,1} + c_5 W^{3,seed}_{\D+1,\ell-1,1} \nn \\ & + c_6 W^{3,seed}_{\D+2,\ell,1} \,. \label{t_channel:Bt2Long}
}
The ansatz for $\Gm_{3,\tilde\Om}^{4}$ comes from the $m_{p+\bar p}\times F$ and $o_{p+\bar p+1} \times \phi$ OPEs, as in the previous exchanged multiplet. The first OPE can be read by looking at the terms proportional to $\t_2^2$ and taking $\t_1=\tb_1=\tb_2=0$, just as before. Looking at \eqref{PO:t2}, we find that such terms are proportional to $\tb_3$ and $\tb_3^2$, which in turn corresponds to the $\Qm$ and $\Qm^2 \bar \Qm$ descendants of the lowest component of the exchanged long multiplet, thus
\eq{
	m_{p+\bar p}\times F \sim G_{\D+1/2}^{\ell} + H_{\D+3/2}^{\ell+1}\,. \label{mF:OPE_2}
}
The $o_{p+\bar p+1} \times \phi$ OPE is found by looking at the term proportional to $\t_1^2$ while setting $\tb_1=\t_2=\tb_2=0$, which yields the same form as in \eqref{mF:OPE_2}:
\eq{
	o_{p+\bar p+1} \times \phi \sim G_{\D+1/2}^{\ell} + H_{\D+3/2}^{\ell+1}\,. \label{oP:OPE_2}
}
Therefore, our ansatz for $\Gm_{3,\tilde\Om}^{4}$ is
\eq{
	-\frac{1}{2}a_1 \frac{1}{u^{1/2}}\Gm_{3,\tilde\Om}^{4}=c_7 g_{\D+1/2,\ell}^{p+\bar p -q-1,q-p-\bar p -1}+c_8 g_{\D+3/2,\ell+1}^{p+\bar p -q-1,q-p-\bar p -1}\,. \label{t_channel:Ct2Long}
}

Plugging \eqref{t_channel:At2Long}, \eqref{t_channel:Bt2Long} and \eqref{t_channel:Ct2Long} into \eqref{OOPPb:t_channel} and \eqref{OOPPb:t_channel_1} why can find the $c_i$ coefficients:
\seq{\label{tc2:02Sol}}{
	c_1 ={}& -\frac{(2 \Delta -3) (\ell +2) (-2 \Delta +2 {p}+2 {\bar p}-2 q+2 \ell +5)^2 (2 \Delta -2 {p}+2 {\bar p}-2 q-2 \ell -1)}{4 (2 \Delta -1) (\ell +1) (2 \Delta -2 \ell -5) (2 \Delta -2 \ell -3) (2 \Delta +2 {p}-2 {\bar p}+2 q-2 \ell -9)}\, \\ 
	c_2 ={}& -\ri a_0 (2 \Delta -2 {p}+2 {\bar p}-2 q+2 \ell +3)\, \\ 
	c_3 ={}& -\frac{\ri a_0 (2 \Delta +2 {p}+2 {\bar p}-2 q+2 \ell +3) (2 \Delta -2 {p}-2 {\bar p}+2 q+2 \ell +3)}{4 (2 \Delta +2 \ell +1) (2 \Delta +2 \ell +3)} \nn \\ & \qquad \times  (2 \Delta +2 {p}-2 {\bar p}+2 q+2 \ell -1)\, \\ 
	c_4 ={}& \frac{\ri \,a_0(2 \D-2 {p}-2 {\bar p}+2 q-2 \ell -5) (4 \D \ell +6 \D-2 {p}+2 {\bar p}-2 q-4 \ell -3)}{(2 \D-3) (2 \D-1) (\ell +1) (\ell +2)} \nn \\
			& \times\left[\frac{1}{4}(-1 + 2 \D - 2 p + 2 {\bar p} - 2 q - 2 \ell) -(p-1)\frac{4 \D \ell +6 \D-2 {p}+2 {\bar p}-2 q-4 \ell -3}{2 \D+2 {p}-2 {\bar p}+2 q-2 \ell -9} \right]\,, \\
	c_5 ={}& \frac{(2\D-1)\ell}{(1+\ell)(2\D-3)}c_1\,c_2\, \\ 
	c_6 ={}& \frac{(1+2\D)(1+\ell)}{(2+\ell)(2\D-1)}c_1\,c_3\, \\ 
	c_7 ={}& \frac{a_1 ({p}-1) (\ell +2) (-2 \Delta +2 {p}-2 {\bar p}+2 q-2 \ell -3) (-2 \Delta +2 {p}+2 {\bar p}-2 q+2 \ell +5)}{(\ell +1) (2 \Delta +2 {p}-2 {\bar p}+2 q-2 \ell -9)}\, \\ 
	c_8 ={}& \frac{a_1 (2 \Delta -3) ({p}-1) (2 \Delta -2 {p}-2 {\bar p}+2 q-2 \ell -5) (2 \Delta +2 {p}+2 {\bar p}-2 q+2 \ell +3) }{4 (2 \Delta -1) (2 \Delta +2 \ell +1) (2 \Delta +2 \ell +3) } \nn \\
			& \qquad \times \frac{(2 \Delta -2 {p}-2 {\bar p}+2 q+2 \ell +3) (2 \Delta +2 {p}-2 {\bar p}+2 q+2 \ell -1)}{(2 \Delta +2 {p}-2 {\bar p}+2 q-2 \ell -9)}
}

Finally, for an $\tilde\Om_{( \left(\D+p-{\bar p}+q-3/2\right)/2,\left(\D-p+{\bar p}-q+3/2\right)/2)}^{(\ell,\ell+1)}$ our ansatz for $\Gm_{0,\tilde\Om}^{4}$, $\Gm_{1,\tilde\Om}^{4}$ and $\Gm_{2,\tilde\Om}^{4}$ are, see \eqref{t_channel:At3}  and \eqref{t_channel:Bt3},
\eq{
	\Gm_{0,\tilde\Om}^{4} ={}& g_{\D+1/2,\ell}^{2-q,q-2} + c_1 g_{\D+3/2,\ell+1}^{2-q,q-2} \label{t_channel:At3Long}
}
and
\eq{
	4 a_0 \left[\left( \frac{\Gm_{1,\tilde\Om}^{4}}{u^{1/2}} +\frac{u^{1/2}}{v} \Gm_{2,\tilde\Om}^{4}\right) {\mathbb T}^4_{\a\ad} + \frac{u^{1/2}}{v} \Gm_{2,\tilde\Om}^{4}\, {\mathbb T}^2_{\a\ad} \right] ={}& c_2 W^{3,dual\,seed}_{\D,\ell,1} + c_3 W^{3,dual\,seed}_{\D+1,\ell+1,1} \nn \\ & + c_4 W^{3,seed}_{\D+1,\ell,1} + c_5 W^{3,dual\,seed}_{\D+1,\ell-1,1} \nn \\ & + c_6 W^{3,dual\,seed}_{\D+2,\ell,1} \,. \label{t_channel:Bt3Long}
}

Finally, the $m_{p+\bar p}\times F$ and $o_{p+\bar p+1} \times \phi$ OPEs are the $\Qm$ and $\Qm^2 \bar \Qm$ descendants of the lowest component of the exchanged long multiplet, which can be seen by comparing the \eqref{PO:t2} and \eqref{PO:t3} correlators. Thus,
\eq{
	m_{p+\bar p}\times F \sim G_{\D+1/2}^{\ell+1} + H_{\D+3/2}^{\ell}\,, \label{mF:OPE_3}\\
	o_{p+\bar p+1} \times \phi \sim G_{\D+1/2}^{\ell} + H_{\D+3/2}^{\ell+1}\,. \label{oP:OPE_3}
}

Therefore, our ansatz for $\Gm_{3,\tilde\Om}^{4}$ is
\eq{
	-\frac{1}{2}a_1 \frac{1}{u^{1/2}}\Gm_{3,\tilde\Om}^{4}=c_7 g_{\D+1/2,\ell+1}^{p+\bar p -q-1,q-p-\bar p -1}+c_8 g_{\D+3/2,\ell}^{p+\bar p -q-1,q-p-\bar p -1}\,. \label{t_channel:Ct3Long}
}

The solutions are
\seq{\label{tc2:03Sol}}{
	c_1 ={}&  -\frac{(2 \D-3) (\ell +1) (2 \D-2 {p}+2 {\bar p}-2 q+2 \ell +5) (2 \D-2 {p}-2 {\bar p}+2 q+2 \ell +1)^2}{4 (2 \D-1) (\ell +2) (2 \D+2 \ell +1) (2 \D+2 \ell +3) (2 \D+2 {p}-2 {\bar p}+2 q+2 \ell -3)}\, \\ 
	c_2 ={}&  -\frac{\ri a_0 (\ell +1) (-2 \D+2 {p}-2 {\bar p}+2 q+2 \ell +3)}{\ell +2}\, \\ 
	c_3 ={}& \frac{(2 \D-1) (\ell +2)}{(2 \D-3) (\ell +1)}c_1\,c_2 \nn \\
		&\qquad \times \frac{(2 \D-2 {p}-2 {\bar p}+2 q+2 \ell +1)^2}{(2 \D+2 {p}-2 {\bar p}+2 q+2 \ell -3)}\, \\ 
	c_4 ={}& \frac{\ri a_0(2 \D-2 {p}-2 {\bar p}+2 q+2 \ell +1) (4 \D \ell +6 \D+2 {p}-2 {\bar p}+2 q-4 \ell -9)}{(2 \D-3) (2 \D-1) (\ell +2)^2}  \nn \\
		& \times \left[\frac{1}{4}(2 \D-2 {p}+2 {\bar p}-2 q+2 \ell +5) + \frac{(p-1)(4 \D \ell +6 \D+2 {p}-2 {\bar p}+2 q-4 \ell -9)}{2 \D+2 {p}-2 {\bar p}+2 q+2 \ell -3} \right]\, \\ 
	c_5 ={}&  \frac{\ri a_0 \ell  (2 \D+2 {p}+2 {\bar p}-2 q-2 \ell -3) (2 \D-2 {p}-2 {\bar p}+2 q-2 \ell -3) }{4 (\ell +1) (2 \D-2 \ell -5) (2 \D-2 \ell -3)} \nn \\ & \qquad \times (2 \D+2 {p}-2 {\bar p}+2 q-2 \ell -7)\, \\ 
	c_6 ={}&  \frac{(2 \D+1) (\ell +1)}{(2 \D-1) \ell } c_1\,c_5\, \\ 
	c_7 ={}& \frac{a_1 ({p}-1) (\ell +1) (-2 \D+2 {p}+2 {\bar p}-2 q-2 \ell -1) (-2 \D+2 {p}-2 {\bar p}+2 q+2 \ell +3)}{(\ell +2) (2 \D+2 {p}-2 {\bar p}+2 q+2 \ell -3)} \, \\ 
	c_8 ={}&  \frac{a_1 (2 \D-3) ({p}-1) (2 \D+2 {p}+2 {\bar p}-2 q-2 \ell -3) (2 \D-2 {p}-2 {\bar p}+2 q-2 \ell -3) }{4 (2 \D-1) (2 \D-2 \ell -5) (2 \D-2 \ell -3) } \nn \\
		& \qquad \times \frac{(2 \D+2 {p}-2 {\bar p}+2 q-2 \ell -7) (2 \D-2 {p}-2 {\bar p}+2 q+2 \ell +1)}{(2 \D+2 {p}-2 {\bar p}+2 q+2 \ell -3)}
}

It is easy to see that in the $p=\bar p=1$ limit, \eqref{tc2:01Sol}, \eqref{tc2:02Sol} and \eqref{tc2:03Sol} we recover \eqref{tc:01Sol}, \eqref{tc:02Sol} and \eqref{tc:03Sol}, respectively. We also recover the results of \cite{Li:2017ddj}, where only the coefficients appearing in $\Gm_{0,\tilde\Om}^{4}$ where obtained.

\section{Conclusions}\label{conclusions}

In this article we have initiated a systematic way of computing the superconformal blocks for general four dimensional $\Nm=1$ SCFTs using the super Casimir operator. After constructing the most general four-point function with total vanishing $U(1)_R$ charge, we were able to compute and solve the super Casimir equations for correlators with one chiral and one anti-chiral field. Our results are in agreement with previous results involving only the superconformal blocks coming from the lowest component of the multiplets. 

There are several interesting generalization of the present work. For example, we can study correlators with non-trivial $\Cm_i$ functions. This will allow us to study the $\langle \Jm\, \Jm\, \Jm \, \Jm\rangle$ four-point function \eqref{JJJJ:4pf}. The three-point function of two currents have, among other multiplets, an $\Om_\D ^{(\ell,\ell+2)}$ and an $\Om_\D ^{(\ell+2,\ell)}$ long multiplets, whose superblocks are identical if one only studies the superprimaries \cite{Berkooz:2014yda}. A complete analysis of the four point function will distinguish the contributions coming from the descendants. 
Finally, including the $\Dm_i$ and $\Em_1$ functions will allow us to study a four-point function solely composed by long multiplets. Another interesting generalization is to include non-trivial spin representations in our four point function. This will be a first step to write the superconformal blocks of the R-current multiplet \cite{Manenti:2018xns}. Including long multiplets with non-zero spin might also impose constraints to the spectrum of the theory when including chiral fields.

Furthermore, most of our techniques are readily generalized to both extended supersymmetries and general dimensions. For example, in the case of extended supersymmetry, the four point functions $\langle \Phi \bar \Phi \Om_1 \Om_2 \rangle$ will depend only on $\Bm_i$ and $\Cm_i$ type of functions for any four dimensional SCFT. This is due to the strong restriction imposed by the shortening conditions which will only be satisfied for the $\Am_1$, $\Bm_i$ and $\Cm_i$ functions. This statement will also be true when studying the four-point function for general dimensions, after taking care of the spinor indices for general dimensions, see \cite{Bobev:2015jxa,Bobev:2017jhk} for a working example of this. We also point out that the stress-tensor multiplet in $\Nm=2$ theories, $\Jm_{\Nm=2}$, satisfies similar constrains as the current multiplet here studied \cite{Kuzenko:1999pi,Liendo:2015ofa,Li:2018mdl}, thus, we argue that the techniques presented in this article are easily generalized to include this multiplet in four-point functions. As has been shown in previous articles, four point functions including at least two stress-tensor multiplets yield strong bounds to the central charge and the flavor central charge \cite{Liendo:2015ofa,Lemos:2015orc}. Recently the superconformal blocks for the lowest component of the stress-tensor were found in \cite{Li:2018mdl}. We propose to use the super Casimir in order to obtain all the superblocks for this correlator and also for mixed systems including the moment map operators.

Finally, as we mentioned, it is possible to write $\Gm_{0,\Om}^{0}$ \eqref{scpw:chirals} as one superblock instead of a sum of several conformal blocks \cite{Fitzpatrick:2012yx}. In order to do so, it necessary to study the asymptotic behavior of the superconformal partial waves using, for example, the supershadow formalism \cite{Fitzpatrick:2012yx,Khandker:2014mpa}. \footnote{A similar analysis was used in order to make an ansatz for the seed blocks \cite{Echeverri:2016dun} using the shadow formalism \cite{Ferrara:1972xe,Ferrara:1972uq,Ferrara:1972ay,Ferrara:1973vz,SimmonsDuffin:2012uy}} To find such simple expression for the $\Gm_{i,\Om}^{1,2,3,4}$ superconformal blocks would greatly simplify computations, allowing for a more straightforward tool to solve the super Casimir equations.

\section*{Acknowledgments}

I would like to thank P. Liendo and M. Lemos for collaborations at early stages of this work. I would also  like to thank Nikolay Bobev, Edoardo Lauria, Per Sundell and Gerben Venken for useful comments and inspiring discussions. I would especially like to thank S. Retamales, for her comments, suggestions and much more. Finally, I would like to thanks Helena Ram\'irez for pointing out the missing $\Cm_1$ term in \eqref{OOppb:4pf} and \eqref{OOppb:4pf_tchannel}. I.A.R. is supported by FONDECYT grant No. 3170373.




\bibliographystyle{abe}
\bibliography{bibliography}

\end{document}